\documentclass[prb,twocolumn, floatfix]{revtex4}
\usepackage{epsfig}
\usepackage{amssymb}

\newcommand{\rv}{\vec{r}}

\newcommand{\rvp}{\vec{r}\,'}
\newcommand{\kv}{{\vec{k}}}
\newcommand{\kvp}{\vec{k}\,'}
\newcommand{\kvi}{\vec{k}^{(i)}}
\newcommand{\kvj}{\vec{k}^{(j)}}
\newcommand{\kvk}{\vec{k}^{(k)}}
\newcommand{\kvl}{\vec{k}^{(l)}}

\newcommand{\khjt}{\vec{k}^{(j)\dagger}}
\newcommand{\ui}{u^{(i)}}
\newcommand{\uj}{u^{(j)}}
\newcommand{\uk}{u^{(k)}}
\newcommand{\ul}{u^{(l)}}
\newcommand{\ujs}{u^{(j)*}}

\newcommand{\ujz}{u^{(j, 0)}}
\newcommand{\kh}{{\hat{k}}}

\newcommand{\khj}{\hat{k}^{(j)}}
\newcommand{\Rv}{\vec{R}}
\newcommand{\snabla}{\nabla_{\rv}}
\newcommand{\lnabla}{\nabla_{\Rv}}
\newcommand{\uv}{\vec{u}}
\newcommand{\Ln}{\textrm{Ln}\,}
\newcommand{\kj}{k^{(j)}}

\newcommand{\Mmatrix}{\mathbf{M}}
\newcommand{\epsnew}{\varepsilon}
\newcommand{\Anew}{\bar{A}}
\newcommand{\Ajnew}{\bar{A}^{(j)}}
\newcommand{\Ajbarsnew}{\bar{A}^{(\bar{j})*}}
\newcommand{\Ajsnew}{\bar{A}^{(j)*}}
\newcommand{\Rnew}{\bar{R}}
\newcommand{\Xnew}{\bar{X}}
\newcommand{\Boxnew}{\bar{\Box}}
\newcommand{\nablanew}{\bar{\nabla}}

\newcommand{\rotp}[1]{{#1}_+}
\newcommand{\rotm}[1]{{#1}_-}
\newcommand{\epspfc}{\epsilon_{2D}}
\newcommand{\Ajpfc}{A_j}

\newcommand{\khn}[1]{\hat{k}^{(#1)}}
\newcommand{\khnd}[1]{\hat{k}^{(#1)\dagger}}
\newcommand{\intmob}{M}
\newcommand{\tempcoupling}{\phi}
\newcommand{\tempcouplingnew}{\bar{\phi}}
\newcommand{\Knorm}{q_0}
\newcommand{\Kabs}{q}
\newcommand{\Boxpfc}{\tilde{\Box}}
\newcommand{\Tpfc}{\tilde{T}}

\newcommand{\pfcpropfac}{\epspfc}
\newcommand{\Ajpfcnew}{\tilde{A}_j}
\newcommand{\Apfcnew}{\tilde{A}}
\newcommand{\epspfcnew}{\tilde{\epsilon}_{2D}}
\newcommand{\nnabla}{\nabla_{\tilde{R}}}
\newcommand{\Boxpfcnew}{\tilde{\Box}}

\begin{document}

\title{Amplitude equations for polycrystalline materials with interaction between composition and stress}

\author{Robert Spatschek\footnote{Present address: Interdisciplinary Centre for Advanced Materials Simulation, Ruhr-Universit\"at 44801 Bochum, Germany}} \author{Alain Karma}
\affiliation{Physics Department and Center for Interdisciplinary Research on Complex Systems, Northeastern University, Boston, MA 02115, USA}

\begin{abstract}
We investigate the ability of frame-invariant amplitude equations [G. H. Gunaratne, Q. Ouyang, and H. Swinney, Phys. Rev. E {\bf 50}, 2802 (1994)] to describe quantitatively the evolution of polycrystalline microstructures and we extend this approach to include the interaction between composition and stress. Validations for elemental materials include studies of the Asaro-Tiller-Grinfeld morphological instability of a stressed crystal surface, polycrystalline growth from the melt, grain boundary energies over a wide range of misorientation, and grain boundary motion coupled to shear deformation.
Amplitude equations with accelerated strain relaxation in the solid are 
shown to model accurately the Asaro-Tiller-Grinfeld instability. Polycrystalline growth is also well described. However, the survey of grain boundary energies shows that the approach is only valid for a restricted range of misorientations as a direct consequence of an amplitude expansion. This range covers approximately half the complete range allowed by crystal symmetry for some fixed reference set of density waves used in the expansion. Over this range, coupled motion to shear is well described by known geometrical rules and a transition from coupling to sliding motion is also reproduced. Amplitude equations for alloys are derived phenomenologically in a Ginzburg-Landau spirit. Vegard's law is shown to be naturally described by seeking a gauge invariant form of those equations under a transformation that corresponds to a lattice expansion and deviations from Vegard's law can be easily incorporated.  
Those equations realistically describe the dilute alloy limit and 
have the same flexibility as conventional phase-field models for incorporating arbitrary free-energy/composition curves. As a test of this approach,  
we recover  known analytical expressions for open-system elastic constants [F. C. Larch\'e and J. W. Cahn, Acta metall. {\bf 33}, 331 (1985)].  
\end{abstract}

\date{\today}
\maketitle

%%%%%%%%%%%%%%%%%%%%%%%%%%%%%%%%%%%%%%%%%%
\section{Introduction and summary}
\label{intro}

Rapid advances in phase-field modeling over the last two decades have greatly enhanced our ability to model a wide range of complex interfacial patterns in materials \cite{Chen02,Boeetal02,Kar05,Steinbach09,WangLi09}. In mature applications such as dendritic solidification, it has been possible to bridge successfully atomistic and continuum scales by linking molecular dynamics and phase field simulations \cite{Braetal02,Hoyetal03,Haxetal06,Astetal09}. This bridge has relied on the combination of thin-interface asymptotic analyses of phase-field models 
\cite{KarRap98,Kar01,Echetal04} to simulate interface dynamics on experimentally relevant length and time scales, and of new atomistic simulation methods to predict some key parameters for those problems such as the anisotropy of the crystal-melt interface \cite{Astetal09}. 

Despite this progress, simulating the evolution of polycrystalline patterns has remained challenging.
Those patterns have been modeled using multiple phase fields \cite{Chen02,Steinbach09}, each representing a different crystal orientation, or by introducing a scalar order parameter that represents the local crystal orientation \cite{Kobetal00,Waretal03}. At a more microscopic level, the phase-field-crystal (PFC) approach  has emerged as an attractive alternative \cite{Eldetal02,Eldetal04,Steetal06,Beretal06,Eldetal06,Proetal07,Berry08,Mellenthin08,Wu09,Jaaetal09,Adletal10}. The PFC model \cite{Eldetal02,Eldetal04} is a reformulation of the Swift-Hohenberg (SH) model of pattern formation \cite{SwiHoh77} with conserved dynamics and can also be motivated as a simplified version of classical density functional theory (DFT) \cite{Eldetal06}. By modeling directly the crystal density field, it provides a simple frame invariant description of polycrystals and it naturally incorporates defects and elastic interactions. However, resolving the density field on the scale of the 
lattice spacing limits the system sizes that can be studied. Therefore, finding ways to ``coarse grain'' spatially 
the PFC model while retaining its advantages is highly desirable, both for computational purposes and to gain analytical insights into the model properties. 
The amplitude equation approach  \cite{NewWhi69,Seg69}, widely used in a pattern formation context, 
\cite{CroHoh93,Gunetal94,Gra96,Gra98e,MatNoz98,Gra98r} has been 
revived recently as a method to coarse-grain the PFC model \cite{Goletal05,Athetal06,Athetal07,Shi09,Goletal09,ChaGol09,Eldetal09}. Amplitude equations have also been used to study the properties of crystal-melt interfaces \cite{Wu07}. 

\subsection{Frame-invariant amplitude equations}

The amplitude equation approach was pioneered by Newell-Whitehead-Segel (NWS) \cite{NewWhi69,Seg69} to model stripe patterns in Rayleigh-B\'enard convection. While NWS derived an amplitude equation from a multiple scale analysis of the 
Navier-Stokes equations, the same equation can be
derived from the SH model where the pattern is described by some scalar field $\Psi$. For stripes in one dimension, the amplitude equation approach exploits the fact that the pattern is slowly modulated in space close to the onset of instability, with the distance from onset measured by a dimensionless parameter $\epsilon$. Hence $\Psi$ can be written as a sum of plane waves 
\begin{equation}
\Psi=u(X)e^{iq_0x}+u^*(X)e^{-iq_0x},
\end{equation}
where the complex amplitude $u(X)$ varies spatially on the slow scale ``$X=\epsilon^{1/2}q_0x$'' as opposed to the original fast scale $q_0x$, which are both defined here to be dimensionless. In addition, $q_0=2\pi/a$ where  $a$, the wavelength of the stripe pattern, is the analog of the ``lattice spacing''. In this framework, coarse graining consists of deriving an amplitude equation, i.e. an evolution equation for $u$, which can be solved on the slow scale $X$. 
The NWS amplitude equation, which is 
derived formally from a multiple scale analysis that exploits the smallness of $\epsilon$, obeys the gradient dynamics $\dot u\sim -\delta F/\delta u^*$ with the Lyapunov functional 
${\cal F}\sim \int dx \left[|\partial_Xu|^2+f_b(u)\right]$ and a bulk energy term $f_b(u)$ that will be specified later. 

As a consequence of this coarse graining, the NWS amplitude equation is not frame invariant (in contrast to the dynamical equation for $\Psi$) because of the fixed choice of reference axis for the underlying plane waves. To overcome this limitation, Gunaratne, Ouyang, and Swinney (GOS) \cite{Gunetal94} have derived a more general frame-invariant form of the NWS equation in the application of this approach to two-dimensional hexagonal patterns. For a ``crystalline'' pattern, 
$\Psi$ can be written as
\begin{equation} \label{dft::eq1}
\Psi(\rv) = \sum_{j=1}^N \uj (\rv) \exp(i\kvj\cdot \rv),\label{Psihex}
\end{equation}
where the sum of plane waves 
is taken here over all principal reciprocal lattice vectors $\kvj$ (with $N=6$ for hexagonal ordering) 
and the complex amplitudes have the 
property that $u^{(j)}=(u^{(l)})^*$ if $\vec{k}^{(j)}=-\vec{k}^{(l)}$, which ensures that $\Psi$ is real, and $|\kvj|=q_0$ for all $j$. The GOS amplitude equations (written here in terms of the fast spatial variable $\vec{r}=x\hat x+y\hat y+z\hat z$, using the hat for normalized vectors) introduce the ``box operator''
\begin{equation}
\Box_j = \khj \cdot \nabla - \frac{i}{2\Knorm} \nabla^2,\label{boxdef}
\end{equation} 
which makes those equations rotationally covariant. 
They have the variational form ($u^{(j)}$ and its complex conjugate $u^{(j)*}$ are considered as independent fields)
\begin{equation}
\partial_t u^{(j)}=-\Gamma \frac{\delta {\cal F}}{\delta u^{(j)*}},\label{notcons}
\end{equation}
where the functional
\begin{equation}
{\cal F}=\int d\vec r \left[\zeta\sum_{j=1}^N|\Box_j u^{(j)}|^2+f_{\rm b}\left(u^{(1)},...,u^{(N)}\right)\right],\label{CalF0}
\end{equation}
and the bulk free-energy density $f_b$ is generally a sum of products of amplitudes for which the sum of
reciprocal lattice vectors form closed polygons up to quartic terms.
More rigorous derivations of frame-invariant amplitude equations using a renormalization group (RG) framework 
have been given for the SH equation \cite{Gra96,Gra98e,MatNoz98,Gra98r} 
and, more recently, the PFC model
\cite{Goletal05,Athetal06,Athetal07,Shi09,Goletal09,ChaGol09}. All those analyses recover
essentially the same form of the GOS amplitude equations with the box operator. The RG
analysis of the PFC model yields amplitude equations with higher order spatial derivatives
\cite{Goletal05,Athetal06} than GOS. However, 
as clarified recently by Chan and Goldenfeld \cite{ChaGol09}, the RG
amplitude equations for PFC can be derived from the same
free-energy functional (\ref{CalF0}) with a form that conserves the volume integral of $\Psi$.
In the present notation, this conserved dynamics is given by
\begin{equation}
\partial_t u^{(j)}=-\Gamma \left(1-2iq_0^{-1}\Box_j\right) \frac{\delta {\cal F}}{\delta u^{(j)}}.\label{cons}
\end{equation}
When expressed in terms of the slow variable
$\vec R=\varepsilon^{1/2} q_0\vec r$, the operator $1-2iq_0^{-1}\Box_j\approx 1$ in the small $\epsilon$ limit of those equations. Hence the non-conserved and conserved dynamics defined by Eqs. (\ref{notcons}) and  (\ref{cons}) are essentially identical in this limit. This limit is physically relevant since $\epsilon$ has been shown to be small in fits of the amplitude equations to actual materials (e.g. pure Fe, see Ref.~\onlinecite{Wu07}). Therefore, in the present work, we use the non-conserved form (\ref{notcons}) that is both accurate and computationally more efficient in this small $\epsilon$ limit. 

Simulations to date support the feasibility of using an amplitude equation approach to simulate polycrystalline microstructural evolution \cite{Goletal05,Athetal06}. Furthermore, they have demonstrated the possibility of significant computational cost saving by using adaptive meshing algorithms \cite{Athetal07}. Despite this progress, the quantitative validity of this approach for modeling polycrystalline patterns is still not fully explored. In addition, for materials application, it would be desirable to extend this approach to alloys. The dual goal of the present work is to explore in more depth both analytically and numerically the quantitative validity of frame-invariant forms of the amplitude equations for simulating polycrystalline pattern evolution and to extend this approach to binary alloys. 

\subsection{Extension to alloys}

One possible approach to model binary alloys is to derive amplitude equations directly from a PFC model with two coupled conserved fields \cite{Eldetal06}. This approach, which was pursued recently by Elder {\it et al.} \cite{Eldetal09}, yields qualitatively similar eutectic phase-diagrams as earlier conventional phase-field models of two-phase growth \cite{Kar94,Eld94}, and has been shown to model complex patterns with defects and elasticity \cite{Eldetal09}. One limitation is that it does not realistically describe the dilute alloy limit and more generally lacks the flexibility of the conventional phase-field approach to model arbitrary liquid and solid free-energy/composition curves \cite{Boeetal02}.

An alternate approach, which is pursued here, is to write down amplitude equations phenomenologically based on physical considerations in the spirit of Ginzburg-Landau theory. For pure materials, this
approach was developed by Shih {\it et al.} \cite{Shietal87} in the framework of DFT to model body-centered-cubic(bcc)/liquid interfaces. Even though there has been subsequent attempts to derive phase field models from DFT \cite{Kha96,Pruessner}, the work of Shih {\it et al.} \cite{Shietal87} was among the first to derive an analytical form for the ``double-well potential'' of the phase field model and the surface energy in the isotropic limit where all density waves have equal amplitudes. In this limit, crystalline order is described by a single scalar variable directly analogous to the phase field.

The amplitude equations of Shih {\it et al.} \cite{Shietal87} were recently revised by Wu {\it et al.} \cite{Wu06} with improved predictions benchmarked against molecular dynamics simulations for pure Fe. 
In this revision, the coefficient of the gradient square term in (\ref{CalF0}) (and quadratic terms in $f_b$)
were related directly to liquid structure factor properties by a small
gradient expansion similar to the one used by Haymet and Oxtoby in a more complete 
DFT study of crystal-melt interfaces \cite{HayOxt81,OxtHay82}. This expansion yields a free-energy
functional of the form (\ref{CalF0}) with $\Box_j\approx \hat{k}^{(j)}\cdot\nabla$, where this truncation
is accurate for the purpose of computing the solid-liquid interfacial free-energy and its anisotropy,
as shown here for the same Fe parameters as in Ref.~\onlinecite{Wu06}.
The coefficients of higher order nonlinearities (cubic or quartic) in $f_b$ were obtained using the
same ansatz as Shih {\it et al.} \cite{Shietal87}. This ansatz holds that all products of amplitudes corresponding to 
polygons with the same number of sides (three or four)  have equal weight \cite{Shietal87}. Wu and Karma \cite{Wu07} have shown that 
this ansatz yields different quartic nonlinearities than in the amplitude equations derived from the PFC model, but that those differences do not alter
significantly solid-liquid interface properties. 

In extending this approach to alloys, two distinct effects of solute addition need to be considered. The first is the coupling between composition and crystalline order.  
This coupling can be introduced phenomenologically by defining a real scalar function  
$\phi(\{u^{(j)}\})$ of the complex amplitudes, which varies between zero in the liquid to one
in the completely ordered solid. Several possible choices of functions that accomplish this goal will be specified later in the paper.
This crystalline order parameter can then be used to interpolate between
the thermodynamic properties of the solid and liquid phases, as in the conventional phase-field approach \cite{Wheetal92}, 
by writing the free-energy density
due to solute addition in the form
\begin{equation}
 f_c(c,T)=\phi(\{u^{(j)}\})f_s(c,T)+\left[1-\phi(\{u^{(j)}\})\right] f_l(c,T),
\end{equation}
where $f_s(c,T)$ and $f_l(c,T)$) are the solid and liquid free-energy/composition curves, respectively, $T$ is the temperature, and $c$ is defined here to be the mole fraction of B in A. To construct an alloy model, $f_c$ needs to be added inside the square brackets in Eq.~(\ref{CalF0}), while making at the same time the bulk free-energy density of the pure material, $f_b$, dependent on temperature.

The second is the coupling between composition and stress. A detailed thermodynamic treatment of the coupling between composition and stress has been given by Larch\'{e} and Cahn (see Ref.~\onlinecite{CahnLarche} with earlier references therein). Solute addition generally modifies the equilibrium  lattice constant of an alloy. 
In its simplest form, the relationship between the lattice constant and concentration at fixed temperature is described by Vegard's law\cite{Vegard}, which is a linear relationship of the form
\begin{equation}  
a=a_0(1-c)+a_1c,\label{Vegardlaw}
\end{equation}
where $a_0$ and $a_1$ are the lattice constants of pure A and pure B, respectively. While this empirical law holds approximately for ionic crystals, metallic alloys show significant deviations from this law. The origin of those deviations has been widely studied theoretically including in the context of classical DFT \cite{Denton}.  

To see how to incorporate the coupling between composition and stress in the amplitude equations, consider a simple one-dimensional crystal (stripe pattern) represented by $\Psi=ue^{iq_0x}+u^*e^{-iq_0x}$. Since considerations of frame invariance are irrelevant in this case, the pure limit of this problem is described by the NWS amplitude equation with a term $~|\partial_x u|^2$ in the free-energy density. A change of lattice constant (wavelength of 
$\Psi$) can be generally represented by the transformation  $u\rightarrow ue^{i p(x)}$. Assuming that $p(x)$ is slowly varying spatially on the lattice scale, this transformation changes locally the lattice constant to $a=2\pi/(q_0+dp/dx)$, which can be approximated by $a\approx a_0 (1-q_0^{-1}dp/dx)$, where $a_0\equiv 2\pi/q_0$. This approximation is valid as long as the change of lattice constant is small $q_0^{-1}|dp/dx| \ll 1$, where
$dp/dx<0$ corresponds to a lattice expansion.   

The transformation $u\rightarrow ue^{i p(x)}$  is directly analogous to the gauge transformation of the quantum mechanical wavefunction generally considered in deriving a gauge invariant form of the Schroedinger equation for a charged particle in an electromagnetic field. Clearly, like the standard Schroedinger equation, the amplitude equation with a term $\sim \partial_x^2u$, derived from the ``kinetic'' part $~|\partial_x u|^2$ of the free-energy density, is not invariant under this transformation. However, this analogy suggests that a gauge-invariant form can be obtained by the substitution $\partial_x\rightarrow \partial_x+i\alpha c$, which transforms the kinetic part into $\sim|(\partial_x+i\alpha c) u|^2$, where $\alpha$ is a coupling constant. The new amplitude equation for $u$ (considered independently from the governing equation for $c$) is now invariant under the transformations $u\rightarrow ue^{i p(x)}$
and $c\rightarrow c-\alpha^{-1} dp/dx$, which yields essentially Vegard's law for the choice of coupling
constant $\alpha=q_0(a_1-a_0)/a_0$.

The generalization to more realistic two- and three-dimensional crystal structures is immediate since a change of lattice parameter corresponds to a change of magnitude of each reciprocal lattice vector $\vec{k^{(j)}}$, and is obtained by the substitution $\Box_j \rightarrow \Box_j+i \alpha c$. Higher order spatial derivatives in the box operator give a negligible contribution since $p(x)$ is slowly varying on the lattice scale. In addition, deviations from Vegard's law can be modeled by the more general transformation $\Box_j \rightarrow \Box_j+i q_0^{-1}h(c)$. This transformation yields the equilibrium lattice constant $a\approx a_0(1+h(c))$. Therefore it can describe an arbitrary relationship between the lattice constant and composition since $h(c)$ can be chosen to be an arbitrary function of $c$.

As an analytical validation of our approach, we derive expressions for the modification of the so-called ``open-systems'' elastic moduli, which agree with the expressions derived by Larch\'{e} and Cahn\cite{CahnLarche}.  The existence of those moduli stems physically from the fact that the crystal strain is generally a function of both composition and stress. In contrast, the composition for an ``open system'' at fixed chemical potential $\mu$, in contact with a reservoir of solute, is only a function of stress. Hence, the strain at fixed $\mu$ can be expressed as a function of stress alone. Consequently, it is generally possible to derive a stress-strain relation at fixed $\mu$ in which the composition has been completely eliminated. This relation yields modified expressions for the elastic moduli with corrections  $\sim \alpha^2$ in the case where Vegard's law applies.  

\subsection{Validations}

There have been several numerical simulations of frame-invariant amplitude equations to date in pure materials \cite{Goletal05,Athetal06,Athetal07,Shi09,Goletal09} and alloys \cite{Eldetal09}. However, quantitative validations of the results have been scarce. Here we carry out quantitative benchmark comparisons with known solutions to test different aspects of the method. We limit those comparisons to pure materials, but the conclusions also pertain to alloys. Validations of the alloy model, beyond the derivation of the open system elastic constants included here, will be presented elsewhere. 

\subsubsection{Asaro-Tiller-Grinfeld instability and strain relaxation}

As a first test, we study the classic Asaro-Tiller-Grinfeld (ATG) \cite{Asaro,Grinfeld} morphological instability of a uniaxially stressed crystal surface, which has been investigated both analytically \cite{Asaro,Grinfeld,Nozieres,Fleck} and numerically \cite{YanSro93,KasMis94,SpeMer94}. The linear stability spectrum (i.e., the exponential amplification rate of sinusoidal perturbations of a solid-liquid interface) is known exactly for this problem and provides a useful quantitative basis for validation. We find that the dynamics defined by Eq. (\ref{notcons}) reproduces qualitatively the instability, but does not predict quantitatively the stability spectrum when the wavelength is much larger than the interface thickness. This is because strain relaxation according to (\ref{notcons}) is diffusive, and hence artificially slow. If strain is not fully relaxed on the time scale that the instability develops, the growth or decay rate of perturbations is altered. This problem was apparently not encountered in a recent PFC simulation study of the same instability \cite{Wu09}. This is possibly due to the shorter wavelengths probed in this study since the PFC with diffusive dynamics suffers the same physical limitation.

To overcome this limitation, we exploit the fact that strain relaxation in the solid can be accelerated in an amplitude equation framework by making the kinetic constant $\Gamma$ a smooth function of the density wave amplitudes. This allows to choose $\Gamma$ much larger in solid than liquid. We find that, with this accelerated strain relaxation   scheme, amplitude equations model accurately the linear regime of the ATG instability. We note that a different acceleration scheme, which has been proposed in the context of the PFC model \cite{Steetal06}, consists of adding inertia to the equations of motion to relax the strain field propagatively, as opposed to diffusively. Although we have not studied this alternative here, it would be interesting to compare the two approaches in the future.

\subsubsection{Polycrystalline growth and grain boundary energies}

The above validation of the method for the ATG instability applies to 
a single crystal. For a fixed set of crystal axes, using 
the box operator or its truncation $\Box_j\approx \hat{k}^{(j)}\cdot\nabla$ gives essentially indistinguishable 
results in the small $\epsilon$ limit. In contrast, for a polycrystal, with different sets of crystal axes pointing in different directions, the full box operator is required to make the amplitude equations frame invariant, as originally proposed by GOS \cite{Gunetal94}. In order to test this frame invariance property, we have studied both the orientation dependence of solid-liquid interfaces, which controls pollycrystalline growth from the melt, and grain boundaries. 

As a strong test of frame invariance for polycrystalline growth, we have computed the solid-liquid gamma plot for two-dimensional hexagonal crystals, i.e. the excess free-energy of the solid-liquid interface, $\gamma_{sl}$, as a function of the angle $\theta$ between the direction normal to the interface and a reference crystal axis, by two methods. First we keep the crystal axes fixed and vary the interface normal direction. Second, we rotate the crystal keeping the normal fixed. Both methods yield identical functions $\gamma_{sl}(\theta)$, thereby validating frame invariance. 

As a strong test of frame invariance for grain boundaries, we have computed the excess free-energy of boundaries, $\gamma_{gb}$, for the whole range of misorientation for symmetric tilt boundaries. These computations were carried out for both two-dimensional hexagonal and three-dimensional bcc ordering. In all cases, we find that frame invariance breaks down for large enough misorientation. 
To illutrate this breakdown, consider symmetric tilt boundaries in two-dimensional hexagonal crystals with the tilt axis normal to the plane of the crystal and $\pm \theta/2$ measuring the rotation angle of each crystal from some fixed axis (e.g. closed packed direction), where $\theta$ denotes now misorientation. The initial increase of $\gamma_{gb}$ with $\theta$ is consistent with a Read-Shockley law \cite{ReaSho50}, as found previously \cite{Goletal05}. However when $\theta$ is increased further, $\gamma_{gb}$ should ultimately vanish by symmetry when $\theta=60^0$ is reached, since the two crystals have the same orientation. Instead, in the amplitude equations, $\gamma_{gb}$ continues to increase and reaches a maximum value at $\theta=60^0$.

\begin{figure}
\begin{center}
\includegraphics[width=6cm]{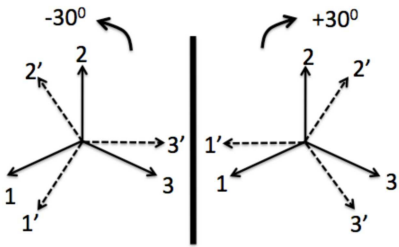}  
\caption{Schematic representation of density waves for two hexagonal crystals rotated by $\pm \theta/2$ where $\theta$ is the misorientation and the thick vertical line represents the grain boundary plane. The unprimed numbers label the fixed set of wavectors $\vec{k}^{(j)}$ with $j=1$ to $3$ and the prime numbers label the wavectors after the rotations. Rotations are produced by spatially oscillating complex amplitudes. Even though the two crystals have the same orientation for the case shown here were $\theta=60^0$, density waves pointing along the same direction are represented by different labels.
}
\label{frameinvariance}
\end{center}
\end{figure}

This unphysical feature originates from the fact that, even though the two crystals have the same orientation when $\theta=60^0$, density waves that point along the same direction have different ``labels'' in each crystal. To see this, 
let us denote by $u^{(1)}$, $u^{(2)}$ and $u^{(3)}$ the amplitudes of density waves with 
wavectors $\vec{k}^{(1)}=-q_0(\sqrt{3}\hat x+\hat y)/2$, $\vec{k}^{(2)}=q_0\hat y$, and $\vec{k}^{(3)}=q_0(\sqrt{3}\hat x-\hat y)/2$, respectively. The three complex conjugates $u^{(1)*}$, $u^{(2)*}$ and $u^{(3)*}$ represent the amplitudes of density waves pointing in opposite directions $-\hat k^{(1)}$,
$-\hat k^{(2)}$, and $-\hat k^{(3)}$, respectively, and the sum of all density waves represents a two-dimensional
hexagonal crystal. For $\theta=0$, both crystals are represented by the same density waves and amplitudes such that $\gamma_{gb}=0$. However,
for $\theta=60^0$, the density wave pointing along the positive $x$-axis ($e^{iq_0x}$) has amplitude $u^{(3)}$ for the crystal rotated by $-30^0$, but amplitude $u^{(1)*}$ for the crystal rotated by $+30^0$, with different permutations for the other wavectors (Fig. \ref{frameinvariance}). Because of the rotations, those amplitudes oscillate rapidly in each crystal with  $u^{(3)}\sim e^{i(-\vec{k}^{(3)}+q_0\hat x)\cdot \vec{r}}$ and $u^{(1)}\sim e^{-i(\vec{k}^{(1)}+q_0\hat x)\cdot \vec{r}}$. The box operator guarantees that those spatial oscillations do not alter the bulk properties of each crystal since the free-energy minimum of (\ref{CalF0}) is frame invariant. However, this frame invariance does not remove the free-energy cost associated with the spatially diffuse $60^0$ rotation of each wavector across the grain boundary.

This breakdown of frame invariance is analyzed in more detail for the simpler case of a smectic crystal in Appendix \ref{smectic}.
This analysis reveals that the grain boundary is ``hidden'' because the reconstructed density field according to Eq. (\ref{dft::eq1}) shows a perfect crystal without any defect.
However, there is nonetheless a high  free energy cost associated with the spatial variation of the phases of the complex amplitudes through the interface. 
This ``hidden boundary'' problem is an intrinsic limitation of the amplitude equations with the free-energy functional (\ref{CalF0}), which extends to other crystal structures. The fact that grain boundaries are hidden may explain why this subtle issue has, to our knowledge, not been explicitly reported or analyzed in the literature so far. Extending the amplitude equation approach to overcome this limitation is an important problem for the future.

For polycrystalline growth, this problem does not occur because density waves do not change direction, but only decrease in amplitude, across the solid-liquid interface. However, a spurious grain boundary energy is created any time that two grains that have the same crystal orientation, but identical density waves labeled with different amplitudes, impinge on each other.

Despite this limitation, we note that the amplitude equation approach is strictly valid for low angle grain boundaries consisting of an array of dislocations in a Read-Shockley picture \cite{ReaSho50}. However, in practice, the description remains approximately valid for misorientations up to about $30^0$ in the example above, or half way between the $\theta=0^0$ and $\theta=60^0$ limits where $\gamma_{gb}$ vanishes by symmetry. Similarly, for symmetric tilt boundaries with a $[001]$ tilt axis in bcc crystals, the regime of approximate validity extends roughly up to about $45^0$, or half way between $\theta=0^0$ and $\theta=90^0$. Also, as demonstrated here, the amplitude equation approach is able to describe the phenomenon of ``grain boundary premelting'', which is associated with the formation of a thin intergranular liquid layer with a width that diverges at the melting point (see Refs.~\onlinecite{Mellenthin08,Hoyetal09} and earlier references therein). Since a continuous film generally only forms for high enough angles where dislocation cores strongly overlap \cite{Mellenthin08}, this indicates that the amplitude equation approach can still capture some interesting grain boundary properties beyond the Read-Shockley picture.
We find that the dependence of the grain boundary energy as function of misorientation agrees quantitatively well with the previous PFC results\cite{Mellenthin08}.

\subsubsection{Grain boundary motion coupled to shear deformation}

As a last validation of the method, we have examined the motion of grain boundaries coupled to a shear deformation. This normal motion is a generic property of grain boundaries that has been widely observed for both low and high angles \cite{Winetal01,Winetal05,Legrosetal08,Giaetal08}. It is generally faster than grain boundary motion associated with diffusional processes and hence can greatly influence the stress-driven evolution of polycrystalline structures. A quantitative understanding of this coupled motion has been obtained from both general theoretical considerations based on geometrical arguments \cite{CahTay04,Cahetal06a,Cahetal06b} and by detailed atomistic simulations \cite{Cahetal06a,Cahetal06b,Mishetal2010}. For low angle boundaries, this motion can be simply understood as the effect of Peach-Koehler forces on individual dislocations, which drives their motion along a direction perpendicular to the grain boundary plane. This motion, however, can still occur for high angle boundaries outside this picture.

For perfectly coupled motion, the velocity perpendicular to the grain boundary plane, $v_\perp$, is proportional to the velocity $v$ of relative translation between the two grains. In this case, the proportionality constant $\beta\equiv v/v_\perp$ depends only on the orientations of the two crystals with a value determined essentially geometrically \cite{CahTay04,Cahetal06a,Cahetal06b}. We find that this perfect coupled motion is well reproduced quantitatively by the amplitude equations for symmetric tilt boundaries in a two-dimensional hexagonal bicrystal where $\beta(\theta)$ is analytically known.

Atomistic simulations have shown that at high homologous temperatures, grain boundaries can become sufficiently disordered to suppress coupled motion, which is superseded by a sliding motion of one grain relative to the other without normal motion \cite{Cahetal06b}. One extreme case of disorder is the formation of a thin intergranular liquid film, associated with the aforementioned grain boundary premelting phenomenon \cite{Mellenthin08,Hoyetal09}. It is clear that the formation of such a film will favor sliding over coupling. We show here that the amplitude equation approach can reproduce this premelting phenomenon and the concominant transition from coupling to sliding when approaching the melting point.

\subsection{Outline}

The rest of this paper is organized as follows. In the next two sections, we construct amplitude equations for elemental materials in a Ginzburg-Landau spirit that parallels the construction of the conventional phase-field model. We discuss first in section II how to derive the ``gradient-square terms'' from classical DFT using a small gradient expansion following previous works \cite{HayOxt81,OxtHay82,Wu06}. We then discuss in section III how to construct the analog of the ``double-well'' potential using the equal weight ansatz or by deriving nonlinearities from the PFC model, which were obtained for bcc in Ref.~\onlinecite{Wu07}. In section IV, still for elemental materials, we derive analytical expressions for the elastic moduli that are generally applicable to different crystal structures and compute their values for parameters of pure Fe determined previously \cite{Wu06,Wu07}. In section V, we then extend the amplitude equations to alloys  and derive analytical expressions for open system elastic constants. The various numerical validations of the results are then presented in section VI following the same order in which they were summarized above. Some technical details have been placed in appendices.

%%%%%%%%%%%%%%%%%%%%%%%%%%%%%%%%%%%%%%%%%%
\section{Gradient-square terms}

We start with the derivation of the quadratic contributions to the free energy density from the classical density functional theory of freezing, following the procedure in Refs.~\onlinecite{HayOxt81,OxtHay82,Wu06} (see also Refs.~\onlinecite{Kha96,Pruessner}). In the liquid phase the time-averaged particle density $n(\rv)$ is spatially constant, whereas it exhibits periodic modulations in the solid, where the atoms have preferential positions.
The free energy is a functional $F=F[n(\rv)]$ of the atomic density, which can be expanded in the form
\begin{equation} \label{dft::eq2}
n(\rv) = n_0 + \delta n(\rv)
\end{equation}
with the homogeneous density $n_0$ of the liquid and
\begin{equation} \label{dft::eq1a}
\Psi(\rv)\equiv \frac{\delta n(\rv)}{n_0} = \sum_{j=1}^N \uj (\rv) \exp(i\kvj\cdot \rv).
\end{equation}
Here, the $\kvj$ are the principal reciprocal lattice vectors;
for bcc, these are
\begin{eqnarray} 
&&[110], [101], [011], [1\bar{1}0], [10\bar{1}], [01\bar{1}], \nonumber \\
&& [\bar{1}\bar{1}0], [\bar{1}0\bar{1}], [0\bar{1}\bar{1}], [\bar{1}10], [\bar{1}01], [0\bar{1}1]. \label{dft::eq1b}
\end{eqnarray}
With the summation $j=1\ldots N$ we write explicitly that we sum over the {\em whole} set of principle reciprocal lattice vectors, thus $N=12$ for bcc.
We point out that we neglect here density differences between the solid and the melt phase.

The goal is to obtain an energy expression in terms of the density wave amplitudes $\uj(\rv)$.
Therefore, we start from the known expression for the free energy change relative to the liquid phase, which changes due to local density variations,
\begin{eqnarray}
\Delta F &=& \frac{k_B T}{2} \int\int d\rv d\rvp \delta n(\rv) \Big[ \frac{\delta(\rv-\rvp)}{n_0} \nonumber \\
&& -C(|\rv-\rvp |) \Big] \delta n(\rvp). \label{dft::eq3}
\end{eqnarray}
Here, $C(r)$ is the direct correlation function of the liquid with Fourier transform
\begin{equation} \label{dft::eq4}
C(\Kabs) = n_0 \int d\rv\, C(r) \exp(-i\kv\cdot\rv)
\end{equation}
with $r=|\rv|$, $\Kabs=|\kv|$.
Therefore, the inverse transformation reads
\begin{equation} \label{dft::eq5}
C(\rv) = \frac{1}{(2\pi)^3 n_0} \int d\kv\, C(\Kabs) \exp(i\kv\cdot\rv).
\end{equation}
We also introduce the liquid structure factor
\begin{equation} \label{dft::eq6}
S(\Kabs) = \frac{1}{1-C(\Kabs)}.
\end{equation}
First, we investigate the integral
\begin{equation} \label{dft::eq7}
I_1 = \int d\rvp \left[ \frac{\delta(\rv - \rvp)}{n_0} - C(|\rv -\rvp|) \right] \delta n(\rvp).
\end{equation}
We assume that each amplitude (density wave envelope) is a slowly varying function, therefore we perform a Taylor series expansion around $\rv$,
\begin{eqnarray}
\uj(\rvp) &=& \uj(\rv) + (\rvp - \rv) \cdot \nabla \uj(\rv) \nonumber \\
&& + \frac{1}{2} (\rvp-\rv)_l (\rvp-\rv)_k \partial_l \partial_k \uj(\rv), \label{dft::eq8}
\end{eqnarray}
where the lower indices in the quadratic term refer to the vector components.
This expansion is inserted into the above integral expression.
The contribution from the $\delta$-function gives readily
\begin{equation} \label{dft::eq9}
\int d\rvp \frac{\delta(\rv - \rvp)}{n_0} \delta n(\rvp) = \frac{1}{n_0} \delta n(\rv).
\end{equation}

Second, the contribution from the $\rvp$ independent term in the series expansion (\ref{dft::eq8}) leads to the integral expression
\begin{eqnarray*}
&& -n_0 \sum_{j=1}^N \uj(\rv) \exp(i\kvj\cdot \rv) \int d\rvp C(|\rv-\rvp |) \times \\
&& \times \exp[i\kvj \cdot(\rvp-\rv)] \\
&=& - \sum_{j=1}^N \uj(\rv) \exp(i\kvj\cdot \rv) C(\Kabs),
\end{eqnarray*}
since $C(\Kabs)=C(\Kabs)^*$ due to the inversion invariance of $C(|\rv |)$ (the star denotes complex conjugation).

Next, we note that the linear term in Eq.~(\ref{dft::eq8}) does not contribute to the remaining part of $I_1$, since we assume that all k-vectors are at the (highest) peak of the structure factor, i.e.~$C'(\Knorm)=0$;
in fact, this term is
\begin{eqnarray*}
-n_0 \sum_{j=1}^N \exp(i\kvj\cdot\rv) \int && d\rvp (\rvp-\rv)\cdot [\nabla \uj(\rv)] \times \\
&& \times C(|\rv-\rvp|)  \exp[i\kvj\cdot (\rvp-\rv)].
\end{eqnarray*}
We therefore inspect the integral (with the notation $\Kabs'=|\kv'|$)
\begin{eqnarray*}
&& n_0 \int d\rvp (\rvp-\rv) C(|\rv-\rvp|)  \exp[i\kv\cdot (\rvp-\rv)] \\
&=&  n_0 \int d\rvp \rvp C(|\rvp|)  \exp[i\kv\cdot \rvp] \\
&=& \frac{1}{(2\pi)^3} \int d\rvp \rvp \int d\kvp C(\Kabs') \exp[i(\kv+\kvp)\cdot\rvp] \\
&=& \frac{-i}{(2\pi)^3} \nabla_\kv \int d\rvp \int d\kvp C(\Kabs') \exp[i(\kv+\kvp)\cdot\rvp] \\
&=& -i \nabla_\kv \int d\kvp C(\Kabs') \delta(\kv + \kvp) \\
&=& -i \nabla_\kv C(\Knorm) = 0.
\end{eqnarray*}

Finally, we look at the term that arises from the quadratic term in the Taylor expansion.
First, we show
\begin{equation}
n_0 \int d\rv C(|\rv |) \exp(i\kv\cdot \rv) \rv_l \rv_j = -C''(\Knorm) \kh_l \kh_j,
\end{equation}
where $\kh_j=\kv_j/\Knorm$ are the normalized components of the vector $\kv$;
the integral is performed on the entire space.
Thus we get
\begin{eqnarray*}
&& n_0 \int d\rv C(|\rv |) \exp(i\kv\cdot \rv) \rv_l \rv_j \\
&=& \frac{1}{(2\pi)^3} \int d\rv \int d\kvp C(\Kabs')\exp[i(\kv+\kvp)\cdot \rv] \rv_l \rv_j \\
&=& -\frac{1}{(2\pi)^3} \frac{\partial^2}{\partial k_l\partial k_j} \int d\rv \int d\kvp C(\Kabs')\exp[i(\kv+\kvp)\cdot \rv] \\
&=& -\frac{\partial^2}{\partial k_l\partial k_j} \int d\kvp C(\Kabs') \delta(\kv+\kvp) \\
&=& -\frac{\partial^2}{\partial k_l\partial k_j} C(\Knorm) \\
&=& -C''(\Knorm) \frac{k_l k_j}{\Knorm^2} = -C''(\Knorm) \kh_l \kh_j
\end{eqnarray*}
since in the last steps $C'(\Knorm)=0$.
With these prerequisites, the last remaining term in $I_1$ becomes
\begin{eqnarray*}
&& -\frac{1}{2} n_0 \sum_{j=1}^N \exp(i\kvj\cdot \rv) \left(\partial_k\partial_l \uj(\rv) \right) \times \\
&& \times \int d\rvp (\rvp-\rv)_k (\rvp-\rv)_l C(|\rvp -\rv|) \exp[i\kvj\cdot(\rvp-\rv)] \\
&=& \frac{1}{2} \sum_{j=1}^N \exp(i\kvj\cdot \rv) \left(\partial_k\partial_l \uj(\rv) \right) C''(\Knorm) \khj_k \khj_l \\
&=& \frac{1}{2} \sum_{j=1}^N \exp(i\kvj\cdot \rv) C''(\Knorm) (\khj\cdot\nabla)^2 \uj(\rv)
\end{eqnarray*}
Altogether, we get
\begin{eqnarray}
I_1 &=& \sum_{j=1}^N \frac{1}{S(\Knorm)} \uj(\rv) \exp(i\kvj\cdot\rv) \\
&+& \frac{1}{2}\sum_{j=1}^N \exp(i\kvj\cdot\rv) C''(\Knorm) (\khj\cdot\nabla)^2 \uj(\rv), \nonumber
\end{eqnarray}
thus the expression for the free energy becomes
\begin{equation}
\Delta F = \frac{k_B T}{2} \int d\rv \delta n(\rv) I_1.
\end{equation}
Now we assume a separation of scales, which means that the scale over which the amplitudes vary is much longer than the atomic spacing $\sim 1/\Knorm$.
Hence we get for a ``slow'' function $g_{s}$
%and here we invoke the multiscale analysis.
%This means, that the spatial variation of the amplitudes $\uj$ is assumed to be slow in comparison to the rapid oscillations of the exponential terms.
%In other words, the decay of the density waves at a solid-liquid interface extends over many atomic units.
%To distinguish between fast and slow variables, we use from now on capital letters for the slow variables.
%Integration becomes then
%\begin{eqnarray}
%\int\limits_{\mathrm{entire\; space}} d\rv &\rightarrow& \sum\limits_{\mathrm{all\; unit\; cells}} \int\limits_{(2\pi/K)^3} d\rv \nonumber \\
%&\rightarrow& \left( \frac{K}{2\pi} \right)^3 \int d\Rv  \int\limits_{(2\pi/K)^3} d\rv.
%\end{eqnarray}
%In particular for the fast exponential oscillations
\begin{equation}
\int d\rv\,\exp(i(\kvj+\kvl)\cdot\rv) g_{s}(\rv) \rightarrow \int d\rv\, \delta_{\kvj+\kvl, 0} g_{s}(\rv).
\end{equation}
Therefore
\begin{eqnarray}
\Delta F &=& \frac{n_0k_B T}{2} \int d\rv \Bigg[ \frac{1}{S(\Knorm)} \sum_{j=1}^N \uj(\rv) {\uj}^*(\rv) \nonumber \\
&+& \frac{1}{2} \sum_{j=1}^N C''(\Knorm) {\uj}^*(\rv) (\khj\cdot\nabla)^2 \uj(\rv) \Bigg],
\end{eqnarray}
where we used that $u^{({l})} = {\ujs}$ if $\kvl=-\kvj$, since the density is real.
We can also integrate the last term by part, assuming that the boundary terms do not contribute for appropriate boundary conditions and obtain finally
\begin{eqnarray}
\Delta F &=& \frac{n_0k_B T}{2} \int d\rv \Bigg[ \frac{1}{S(\Knorm)} \sum_{j=1}^N \uj {\ujs} \nonumber \\
&-& \frac{1}{2} \sum_{j=1}^N C''(\Knorm) |(\khj\cdot\nabla) \uj|^2 \Bigg]
\end{eqnarray}

As pointed out in Refs.~\onlinecite{Gunetal94,Gra96} the operator $\khj\cdot\nabla$ violates the rotational invariance of the functional.
As already mentioned, the proper renormalization is to replace it by the ``box operator''
\begin{equation}
\khj\cdot\nabla \longrightarrow \Box_j = \khj \cdot \nabla - \frac{i}{2\Knorm} \nabla^2.
\end{equation} 
The properties of this operator and the consequences for the model will be discussed in detail in section \ref{box}.

The derivation of the free energy allows to link the phenomenological parameter $\zeta$ in Eq.~(\ref{CalF0}) to physical parameters via the relation $\zeta=-n_0 k_B T C''(q_0)/4$.

%%%%%%%%%%%%%%%%%%%%%%%%%%%%%%%%%%%%%%%%%%
\section{Double well potential}
\label{dw}

We now discuss the derivation of higher order cubic and quartic nonlinear terms. With the addition of those nonlinearities, the bulk free-energy density has the form of the standard ``double-well'' potential of the phase-field model. As already mentioned in Section \ref{intro}, nonlinearities can be obtained in a Ginzburg-landau spirit using the equal weight ansatz of Shih {\it et al.} \cite{Shietal87}, or derived from the PFC model  \cite{Goletal05,Athetal06,Wu07}. Here we give the results of both methods for bcc and 2-d hexagonal cases.

Due to the assumption of a scale separation, orthogonality demands that only higher nonlinearities which form a closed polygon of reciprocal lattice vectors contribute.
The addition of cubic and quartic terms gives therefore the general expression
\begin{eqnarray}
\Delta F&=&\frac{n_0k_BT}{2}\int d\rv \bigg[\frac{1}{S(\Knorm)}\sum_{j=1}^N|\uj|^2 \label{dw::eq1} \\
&& -\frac{C''(\Knorm)}{2}\sum_{j=1}^N|\Box_j \uj|^2 \nonumber \\
& & -a_3\sum_{ijk}\alpha_{ijk}\ui\uj\uk\delta_{0,\kvi+\kvj+\kvk} \nonumber \\
& & +a_4\sum_{ijkl}\alpha_{ijkl}\ui\uj\uk\ul\delta_{0,\kvi+\kvj+\kvk+\kvl} \bigg]. \nonumber
\end{eqnarray}
In the summation over three or four wave vectors in the cubic and quartic terms the summation is normalized such that permuting indices that correspond to equivalent sets of $\kv$-vectors are only counted once. 
Using an equal weight ansatz, one obtains \cite{Wu06}
\begin{equation}
\alpha_{ijk}^{(GL)}=1/8,~\alpha_{ijkl}^{(GL)}=1/27 \label{dw::eq2}
\end{equation}
and
\begin{eqnarray}
a_3 &=& \frac{24}{S(\Knorm)u_s}, \label{dw::eq3} \\
a_4 &=& \frac{12}{S(\Knorm)u_s^2}, \label{dw::eq4}
\end{eqnarray}
where $u_s$ is the amplitude of all density waves in the solid phase \cite{Wu06,Wu07}.
The connection to a conventional phase field model with a double well potential becomes obvious in the isotropic approximation, where all density wave amplitudes are equal, $\uj=u$, and assumed to be real.
Then the local part of the free energy density (\ref{dw::eq1}) becomes
\begin{equation}
\Delta f_{dw} = \frac{n_0 k_B T}{2} \frac{12}{S(\Knorm) u_s^2} u^2 (u-u_s)^2,
\end{equation}
which obviously has energetically equivalent minima for the bulk states $u=0$ and $u=u_s$.

Alternatively, the coefficients of the higher order nonlinearities can be derived using a multiscale analysis of a phase field crystal model, yielding different expressions for the quartic coefficients only \cite{Wu07}:
\begin{equation} \label{dw::eq5}
\alpha_{ijkl}^{(AE)} = \left\{ 
\begin{array}{ll}
\frac{1}{90} & \mbox{only two wave vectors $\kvj$ and $-\kvj$} \\
\frac{4}{90} & \mbox{all other quartic}
\end{array}
\right.
\end{equation}

So far, the minima of the free energy correspond to solid and liquid phases which are energetically equivalent.
A deviation from the melting temperature $T_M$ favors one or the other phase, and this is achieved here by introducing a tilt term of the form
\begin{equation} \label{dw::eq5a}
F_T = \int d\rv L\frac{T-T_M}{T_M} \tempcoupling(\{ \uj\})
\end{equation}
with the latent heat $L$ and a coupling function $\tempcoupling$ (analogous to the standard ``phase field'') that has value $1$ in the solid and $0$ in the liquid;
the function $\tempcoupling$ should be stationary there in order not to shift the bulk states $\uj=0$ and $\uj=u_s$.
A particular choice is
\begin{equation} \label{dw::eq5b}
\tempcoupling(\{ \uj\}) = \frac{1}{N} \sum_{j=1}^N h(|\uj|^2/u_s^2)
\end{equation}
with
\begin{equation} \label{dw::eq5c}
h(x)=x^2(3-2x)
\end{equation}
or 
\begin{equation} \label{dw::eq5d}
h(x)=x(3-2\sqrt{x}).
\end{equation}
For the simulations shown here we used the first choice, Eq. (\ref{dw::eq5c}). Other choices are possible as in the conventional phase-field approach.

For the purpose of numerical implementation, it is useful to rewrite the free energy in a dimensionless version.
We therefore introduce a new small parameter
\begin{equation} \label{dw::eq6}
\epsnew = -\frac{24}{S(\Knorm)C''(\Knorm) \Knorm^2},
\end{equation}
a dimensionless amplitude
\begin{equation} \label{dw::eq7}
\Ajnew = \uj/u_s,
\end{equation}
and a dimensionless (slow) lengthscale
\begin{equation} \label{dw::eq8}
\Xnew = x \epsnew^{1/2} \Knorm.
\end{equation}
We note that $\epsnew$ is defined differently here for bcc than in Ref. \onlinecite{Wu07} to eliminate all PFC parameters from the amplitude equations.
Then the free energy becomes for the phase field crystal model
\begin{eqnarray}
F_{AE} &=& F_0 \int d\Rnew \Bigg[ \sum_{j=1}^{N/2} | \Boxnew_j \Ajnew|^2 + \frac{1}{12} \sum_{j=1}^{N/2} \Ajnew \Ajsnew \nonumber \\
&&+ \frac{1}{90} \Bigg\{  \left( \sum_{j=1}^{N/2} \Ajnew \Ajsnew \right)^2 - \frac{1}{2} \sum_{j=1}^{N/2} |\Ajnew|^4 \nonumber \\
&& + 2\Anew_{110}^* \Anew_{1\bar{1}0}^* \Anew_{101} \Anew_{10\bar{1}} + 2\Anew_{110} \Anew_{1\bar{1}0} \Anew_{101}^* \Anew_{10\bar{1}}^* \nonumber \\
&& + 2\Anew_{1\bar{1}0} \Anew_{011} \Anew_{01\bar{1}} \Anew_{110}^* + 2\Anew_{1\bar{1}0}^* \Anew_{011}^* \Anew_{01\bar{1}}^* \Anew_{110} \nonumber \\
&& + 2\Anew_{01\bar{1}} \Anew_{10\bar{1}}^* \Anew_{101} \Anew_{011}^* + 2\Anew_{01\bar{1}}^* \Anew_{10\bar{1}} \Anew_{101}^* \Anew_{011} \Bigg\} \nonumber \\
&& - \frac{1}{8} \Bigg\{ \Anew_{011}^* \Anew_{101} \Anew_{1\bar{1}0}^* +  \Anew_{011} \Anew_{101}^* \Anew_{1\bar{1}0} \nonumber \\
&& +  \Anew_{011}^* \Anew_{110} \Anew_{10\bar{1}}^* +  \Anew_{011} \Anew_{110}^* \Anew_{10\bar{1}} \nonumber \\
&& +  \Anew_{01\bar{1}}^* \Anew_{110} \Anew_{101}^* +  \Anew_{01\bar{1}} \Anew_{110}^* \Anew_{101} \nonumber \\
&& +  \Anew_{01\bar{1}}^* \Anew_{10\bar{1}} \Anew_{1\bar{1}0}^* +  \Anew_{01\bar{1}} \Anew_{10\bar{1}}^* \Anew_{1\bar{1}0}\Bigg\} \Bigg] \label{dw::eq9}
\end{eqnarray}
with
\begin{equation} \label{dw::eq10}
F_0 = -\frac{n_0 k_B T}{2} C''(\Knorm) \Knorm^{-1} u_s^2 \epsnew^{-1/2}.
\end{equation}
Here, the box operator is defined as
\begin{equation} \label{dw::eq11}
\Boxnew_j =  \khj\cdot\nablanew - \frac{i\epsnew^{1/2}}{2} \nablanew^2 = \Knorm^{-1} \epsnew^{-1/2} \Box_j,
\end{equation}
where the nabla operator $\nablanew$ acts on the variable $\Rnew$.
The amplitudes $\Ajnew$ are defined as functions of the (dimensionless) ``slow'' scale $\Xnew$, which is much larger than the atomic spacing for $\epsnew \ll 1$.
The summation $N/2$ in the expression (\ref{dw::eq9}) expresses that we sum only over the contributions from independent density waves, since $\Ajbarsnew = \Ajnew$.
It means e.g.~for bcc that we sum only over the six principal reciprocal lattice vectors $[110]$, $[101]$, $[011]$, $[1\bar{1}0]$, $[10\bar{1}]$, $[01\bar{1}]$.

Similarly, the thermal tilt (\ref{dw::eq5a}) becomes
\begin{equation}
F_T = \epsnew^{-3/2} \Knorm^{-3} \int d\Rnew\, L \frac{T-T_M}{T_M} \tempcouplingnew(\{\Ajnew\})
\end{equation}
with
\begin{equation}
\tempcouplingnew(\{\Ajnew\}) = \frac{2}{N} \sum_{j=1}^{N/2} h(|\Ajnew|^2).
\end{equation}

The reconstructed density becomes
\begin{eqnarray}
n(\Rnew) &=& n_0 \left[ 1 + u_s \sum_{j=1}^N \Ajnew(\Rnew) \exp\left( \frac{i\khj\cdot\Rnew}{\epsnew^{1/2}} \right) \right] \nonumber \\
&=& n_0 \left[ 1 + \sum_{j=1}^N \uj(\Rnew) \exp\left( \frac{i\khj\cdot\Rnew}{\epsnew^{1/2}} \right) \right]. \label{dw::eq12}
\end{eqnarray}

The dynamical evolution of the nonconserved order parameters is described by
\begin{equation}
\frac{\partial \Ajnew}{\partial t} = -K_j \frac{\delta F}{\delta \Ajsnew}.
\end{equation}
As long as we are interested only in equilibriums properties, the choice $K_j>0$ (or correspondingly $\Gamma$ in Eq.~(\ref{notcons})) is not important. The formulation of the dynamics
will be discussed in detail in Section \ref{Grinfeld}.

Similarly, for a two-dimensional hexagonal model we get the (rescaled) free energy from a phase field crystal model (see Appendix \ref{hex} for details)
\begin{eqnarray}
F_{2D}^{\mathrm{PFC}} &=& \tilde{F}_{2D}^0 \int d\tilde{R} \Bigg\{ \sum_{j=1}^{N/2} \left| \Boxpfcnew_j \Ajpfcnew \right|^2 \nonumber + \frac{1}{6} \sum_{j=1}^{N/2} \Ajpfcnew \Ajpfcnew^* \nonumber \\
&&+ \frac{1}{2} (\Apfcnew_1^{*} \Apfcnew_2^{*} \Apfcnew_3^{*} + \Apfcnew_1 \Apfcnew_2 \Apfcnew_3) \nonumber \\
&& + \frac{1}{15} \left(\sum_{j=1}^{N/2} \Apfcnew_j \Apfcnew_j^{*} \right)^2 - \frac{1}{30} \sum_{j=1}^{N/2} |\Apfcnew_j|^4 \Bigg\}
\end{eqnarray}
where we have $N=6$ principal reciprocal lattice vectors, see Eq.~(\ref{hex::eq5}).
Assuming an equal weight ansatz, we again get slightly different quartic terms:
\begin{eqnarray}
F_{2D}^{\mathrm{GL}} &=& \tilde{F}_{2D}^0 \int d\tilde{R} \Bigg\{ \sum_{j=1}^{N/2} \left| \Boxpfcnew_j \Ajpfcnew \right|^2 \nonumber + \frac{1}{6} \sum_{j=1}^{N/2} \Ajpfcnew \Ajpfcnew^* \nonumber \\
&&+ \frac{1}{2} (\Apfcnew_1^{*} \Apfcnew_2^{*} \Apfcnew_3^{*} + \Apfcnew_1 \Apfcnew_2 \Apfcnew_3) \nonumber \\
&& + \frac{1}{24} \left(\sum_{j=1}^{N/2} \Apfcnew_j \Apfcnew_j^{*} \right)^2 + \frac{1}{24} \sum_{j=1}^{N/2} |\Apfcnew_j|^4 \Bigg\}.
\end{eqnarray}

%%%%%%%%%%%%%%%%%%%%%%%%%%%%%%%%%%%%%%%%%%
\section{Elasticity}
\label{elast}

The amplitude equations naturally incorporate linear elasticity. 
Here we derive an analytical expressions for the elastic constants that are generally applicable to different crystal structures by summing the contributions of different sets of crystal density waves. Those expressions reduce to linear elasticity for two-dimensional hexagonal crystals and yield reasonable predictions of the elastic moduli for bcc. Both hexagons and bcc can be represented by only the principal set of reciprocal lattice vectors. In contrast, for the PFC model of fcc structures \cite{Adletal10}, the present analysis shows that the addition of a second set of density waves is required to obtain a non-vanishing tetragonal shear modulus. Here we restrict our attention to linear behavior  elastic behavior. Nonlinear elasticity has also been shown to be analytically treatable in an amplitude equation framework\cite{ChaGol09}. 

We start from a density field
\begin{equation} \label{elast::eq1}
\delta n_0(\rv) = n_0 \sum_j \ujz(\rv) \exp(i\kvj\cdot \rv)
\end{equation}
of a stress free system.
The amplitude $\ujz(\rv)$ may describe a pure solid or a system consisting both of solid and liquid parts, and we only assume that the amplitude varies slowly in space.
If e.g.~an external stress is applied to the system, the atoms are displaced and take new positions.
In other words, an atom that was previously located at $\rv$ is now at $\rv+\uv(\rv)$, with $\uv$ being the displacement field.
Ignoring density changes due to the strain, this leads to the condition $\delta n(\rv+\uv(\rv))=\delta n_0(\rv)$.
Since we assumed that the amplitudes themselves are slowly varying functions, we can assume $\ujz(\rv+\uv)\approx \ujz(\rv)$.
Then we obtain for the density field of the stressed sample
\begin{equation} \label{elast::eq2}
\delta n(\rv) = n_0\sum_j \ujz(\rv) \exp(-i\kvj\cdot \uv(\rv)) \exp(i\kvj\cdot\rv).
\end{equation}
This means that the material is now described by new density wave amplitudes
\begin{equation} \label{elast::eq3}
\uj(\rv) = \ujz(\rv) \exp(-i\kvj\cdot \uv(\rv)),
\end{equation}
which are oscillating functions (in a solid), and the wavelength of the oscillation is shorter for higher strains.

The principal reciprocal lattice vectors obey the following orthogonality relation \cite{Pruessner}
\begin{equation} \label{elast::eq4}
\sum_{j=1}^N \kj_l \kj_m = \frac{N}{d} \Knorm^2 \delta_{lm},
\end{equation}
where $N$ is the number of principal reciprocal lattice vectors, $d$ the spatial dimension and $\Knorm=|\kvj|$, which is equal for all principal reciprocal lattice vectors.
Here, the lower index denotes the component of a vector.
It is straightforward to check that this relation holds for the two present cases of interest, the three-dimensional bcc lattice, Eq.~(\ref{dft::eq1b}) and the two-dimensional hexagonal lattice, Eq.~(\ref{hex::eq5}).

We can therefore extract the displacement field from the amplitudes by taking the (complex) logarithm
\begin{equation} \label{elast::eq5}
\Ln \frac{\uj(\rv)}{|\uj(\rv)|} = -i\kvj\cdot \uv(\rv) + 2\pi i m(\rv) + i\phi_0.
\end{equation}
Here, we used the symbolic notation of an integer ``winding number'' $m(\rv)$ which stems from the fact that the imaginary part of the complex logarithm,
\begin{equation} \label{elast::eq6}
\Ln z = \ln |z| + i \arg z,
\end{equation}
is defined only up to arbitrary shifts by $2\pi i$, which can be absorbed in the complex phase $\arg z$.
Since we are mainly interested in derivatives of the above expression, this additional integer term usually disappears anyway.
It becomes only relevant if defects are present in the system.
%, and we will study such situations below.
The last term $\phi_0$ is the phase of the slow amplitude $\ujz$, which may result from a rigid body translation, and which is therefore constant.
Another possibility would be a rigid body rotation, and then $\phi_0$ is not constant.
However, this term does not contribute to the final result and we therefore ignore it here.
%A simple perspective is to assume that the amplitudes are initially pure real.

Differentiation and summation over all fields gives, together with Eq.~(\ref{elast::eq4}),
\begin{eqnarray*}
\sum_j^N \kj_l \partial_m \Ln \frac{\uj(\rv)}{|\uj(\rv)|} &=& -i (\partial_m u_n(\rv)) \sum_j^N \kj_l \kj_n \\
&=& -i\frac{N}{d}\Knorm^2 \partial_m u_l(\rv).
\end{eqnarray*}
We therefore arrive at the following expression for the (infinitesimal) strain tensor $\epsilon_{lm} = (\partial_l u_m + \partial_m u_l)/2$:
\begin{equation} \label{elast::eq7}
\epsilon_{lm} = \frac{i\,d}{2N\Knorm^2} \sum_j (\kj_l\partial_m + \kj_m\partial_l) \Ln \frac{\uj(\rv)}{|\uj(\rv)|}.
\end{equation}
By straightforward algebraic manipulations we obtain the alternative representation
\begin{equation} \label{elast::eq8}
\epsilon_{lm} = -\frac{d}{2N \Knorm^2} \Im \left( \sum_j^N \frac{1}{\uj} (\kj_l\partial_m + \kj_m\partial_l) \uj \right),
\end{equation}
where $\Im(\cdot)$ denotes the imaginary part.

In the expression for the free energy, only the gradient term changes if a solid phase is displaced.
The local terms remain unchanged, since by construction they consist of density waves amplitude products with closed polygons of principal reciprocal lattice vectors.
Therefore, in any product like $u^{(1)} u^{(2)} u^{(3)}$ (for the hexagonal system, see Appendix \ref{hex} for details), we have
\begin{eqnarray*}
u^{(1)} u^{(2)} u^{(3)} &=& u^{(1, 0)} u^{(2, 0)} u^{(3, 0)} \times \\
&& \times \exp[-i(\kv^{(1)}+\kv^{(2)}+\kv^{(3)})\cdot \uv(\rv)]  \\
&=& u^{(1, 0)} u^{(2, 0)} u^{(3, 0)},
\end{eqnarray*}
and it is therefore sufficient to inspect the gradient terms.

%{\bf Now probably the $\epsilon$ scaling is not correct, I should check that later.}
The ``kinetic'' part of the free energy is
\begin{equation} \label{elast::eq9}
F_k = -\frac{1}{2}\frac{n_0 k_B T}{2} C''(\Knorm) \int d\rv \sum_j^N | (\khj\cdot\nabla) \uj|^2.
\end{equation}
Notice that we dropped here the correction term from the box operator.
The reason is that this term gives only a higher order correction, and for elastic deformations (which are always assumed to be long-wave distortions), the additional term is negligible.
Let us assume for simplicity that the original amplitudes $\ujz$ are real.
Then we get
\begin{equation} \label{elast::eq10}
|(\khj\cdot\nabla) \uj|^2 = \left( \khj_l (\partial_l \ujz) \right)^2 + \left(\khj_l \kj_m \ujz \partial_l u_m \right)^2.
\end{equation}
The first term is the usual interfacial energy (the same as for the undeformed state), and the second term, which is quadratic in the distortions, the elastic energy.
Notice that we do not get a term that is linear in the strain, and therefore we do not have a surface stress term in the model.

We get explicitly for the elastic term
\begin{equation} \label{elast::eq11}
f_{el} = -\frac{1}{2} \frac{n_0k_B T}{2} \frac{C''(\Knorm)}{\Knorm^2}  \sum_{j=1}^N \kj_l \kj_m \kj_\alpha \kj_\beta {\ujz}^2 \epsilon_{lm} \epsilon_{\alpha\beta}.
\end{equation}
We can therefore identify the elastic constants
\begin{equation} \label{elast::eq12}
c_{lm\alpha\beta} = -\frac{n_0k_B T}{2} \frac{C''(\Knorm)}{\Knorm^2} \sum_{j=1}^N \kj_l \kj_m \kj_\alpha \kj_\beta {\ujz}^2,
\end{equation}
which are obviously invariant under pairwise permutations of indices.
Here we used 
\begin{equation} \label{elast::eq13}
\sigma_{ij} = c_{ijkl} \epsilon_{kl}
\end{equation}
and the usual expression for the elastic free energy density,
\begin{equation} \label{elast::eq14}
f_{el} = \frac{1}{2} \sigma_{ij}\epsilon_{ij}.
\end{equation}
Notice that the expression for the elastic constants automatically reflects the correct crystallographic symmetries.

First, we note that in the ``liquid'' phase all elastic constants become zero;
this is a consequence of the fact that we skipped contributions from density changes.

For the bcc case with cubic symmetry we obtain
\begin{equation} \label{elast::eq15}
c_{ijkl} = -\frac{n_0k_B T}{2} C''(\Knorm)\Knorm^2 u_s^2 \times
\left\{
\begin{array}{cc}
2 & \mbox{if } i=j=k=l \\
1 & \mbox{two distinct index pairs} \\
0 & \mbox{else}
\end{array}
\right. 
\end{equation}
We can calculate explicitly the predicted values for the elastic constants for bcc iron at the melting point.
Using the parameters given in Ref.~\onlinecite{Wu07}, which are summarized in table \ref{elast::table1}, we obtain (in Voigt notation)
\begin{table*}
\caption{Parameters for bcc iron, see Ref.~\onlinecite{Wu06} for details, as obtained from MD simulations \cite{bcciron}.
%The Thermocalc value for the latent heat is $L=1.47\cdot 10^9\, \mathrm{J} \mathrm{m}^{-3}$, the number that I used originally is $1.968\cdot 10^9\, \mathrm{J} \mathrm{m}^{-3}$. The number in the table is what I used, so maybe I should rescale some of the temperatures. At the moment this is only relevant for the temperature value in Fig. 10, but I will have to remember it for the short range force paper...}
}
\label{elast::table1}
\begin{tabular}{ccccccc}
\hline\hline
$\Knorm$ & $S(\Knorm)$ & $C''(\Knorm)$ & $u_s$ & $n_0$ & $T_M$ & $L$ \\
\hline
$2.985\cdot 10^{10} \mathrm{m}^{-1}$ & $3.01$ & $-10.4\cdot10^{-20} \mathrm{m}^2$ & $0.72$ & $0.0765\cdot10^{30} \mathrm{m}^{-3}$ & $1773\, \mathrm{K}$ & $1.968\cdot 10^9\, \mathrm{J} \mathrm{m}^{-3}$ \\
\hline\hline
\end{tabular}
\end{table*}
\begin{eqnarray}
C_{11} &=& C_{22} = C_{33} = 90\, \mathrm{GPa}  \label{elast::eq16}\\
C_{12} &=& C_{23} = C_{44} = 45\, \mathrm{GPa}. \label{elast::eq17}
\end{eqnarray}
Given that the theory only includes one set of density waves, those values are reasonably good (see Ref. \onlinecite{Adletal10} for a quantitative comparison with MD results).

The two-mode PFC model for fcc structures couples two different sets of crystal density waves corresponding to $\langle 111\rangle$ and $\langle 200\rangle$ reciprocal lattice vectors\cite{Adletal10}. Using the analytical expression for the elastic constants, Eq. (\ref{elast::eq12}), it is straightforward to work out that the $\langle 111\rangle$ set yields equal elastic moduli $C_{12}=C_{22}=C_{44}$, and hence vanishing tetragonal modulus $(C_{11}-C_{12})/2$. However this unphysical feature is cured by the addition of the $\langle 200\rangle$ set that brings an additional finite contribution to $C_{11}$ and vanishing contributions to $C_{12}$ and $C_{44}$. Therefore, with both the $\langle 111\rangle$ and $\langle 200\rangle$ sets present, $C_{11}>C_{12}$ and $C_{12}=C_{44}$, such that the tetragonal shear modulus is finite as physically desired. Furthermore, the expressions for the elastic moduli predicted by Eq. (\ref{elast::eq12}) agree with those derived in Ref. \onlinecite{Adletal10} by a brute force calculation of quadratic contributions to the PFC two-mode free-energy functional for different lattice distortions.   

Next, from the equilibrium conditions we know that the free energy $F$ has to be minimized with respect to all degrees of freedom.
For fixed interface position this implies that in particular $F$ has to minimized with respect to the elastic displacements, which appear only in the elastic contribution $f_{el}$.
Therefore, we get $\delta F/\delta u_i = 0$ and consequently
\begin{equation} \label{elast::eq18}
\frac{\partial\sigma_{ij}}{\partial x_j} = 0,
\end{equation}
which are the usual elastic equations.

Finally, we note that the model allows for deformations that are not contained in the standard theory of linear elasticity.
The reason is that the displacement vector has only $d$ components, but we have have $N/2$ independent amplitudes.
This means that not all possible amplitudes can be represented in the form (\ref{elast::eq3}) with real amplitudes $\ujz$ of the ``undeformed'' crystal.
The remaining $N/2-d$ degrees of freedom correspond to atomic ``shuffles'' i.e. rearrangements within each unit cell.
%Since they cost unnecessary free energy, they will usually not be excited.

%%%%%%%%%%%%%%%%%%%%%%%%%%%%%%%%%%%%%%%%%%
\section{Alloys and Vegard's law}
\label{alloy}

We now consider the extension of the amplitude equations to binary alloys. As discussed in section I, the present extension has the advantage that it can interpolate between the thermodynamic properties of the solid and liquid phases, which can be described in principle by arbitrary free-energy/composition curves.
For concreteness, we consider here a binary alloy in the dilute regime.

The impurities are introduced variationally using a new free energy term
\begin{equation} \label{alloy::eq1}
F_c +F_T = \int d\rv \left[ \frac{R T_M}{v_0} \left( c\ln c - c \right) + \tempcoupling \Delta\epsilon\, c + L \frac{T-T_M}{T_M} \tempcoupling \right],
\end{equation} 
which includes the previous temperature coupling and the phase field $\tempcoupling$ as defined in Eq.~(\ref{dw::eq5b}).
% BELOW IS THE OLD DEFINITION
%The ``phase field'' is defined as 
%\begin{equation}
%\phi = \chi([|A_1|^2 + |A_2|^2 + |A_3|^2]/3 A_s^2)
%\end{equation}
%where $A_s$ is the absolute value of the amplitudes in the solid.
%The function $\chi$ interpolates between $0$ in the liquid (all amplitudes zero) and $1$ in the solid;
%a possible choice is $\chi(z) = z(3-2\sqrt{z})$, which has a stationary point both in the liquid and the solid, $\uj=u_s$.
It contains the molar volume $v_0$ and the ideal gas constant $R$.

The evolution equation for the amplitudes is the same as before,
\begin{equation} \label{alloy::eq2}
\frac{\partial \uj}{\partial t} \sim - \frac{\delta F}{\delta \ujs},
\end{equation}
where we get now the additional term
\begin{equation} \label{alloy::eq3}
\frac{\delta (F_c+F_T)}{\delta \ujs} = \left( L \frac{T-T_M}{T_M} + \Delta\epsilon\, c \right) h'(|\uj/u_s|^2) \frac{\uj}{N u_s^2}.
\end{equation}
The diffusion equation follows from
\begin{equation} \label{alloy::eq4}
\frac{\partial c}{\partial t} = \nabla\cdot \left[ D \frac{v_0}{R T_M} c \nabla \frac{\delta F}{\delta c} \right]
\end{equation}
with a diffusion coefficient that can be different in solid and liquid, and gives therefore
\begin{equation} \label{alloy::eq5}
\frac{\partial c}{\partial t} = \nabla\cdot \left[ D \nabla c - D b c \nabla \phi \right]
\end{equation}
with
\begin{equation} \label{alloy::eq6}
b = - \frac{v_0\Delta\epsilon}{R T_M}.
\end{equation}

These equations describe in equilibrium a phase diagram with straight solidus and liquidus lines.
The partition coefficient is given by
\begin{equation} \label{alloy::eq7}
k = \exp(b),
\end{equation}
and the liquidus slope is
\begin{equation} \label{alloy::eq8}
m = -\frac{R T_M^2}{v_0 L} (1-k).
\end{equation}

Additionally, we can take into account that the equilibrium lattice constant changes with impurity concentration.
For a linear dependence (Vegard's law\cite{Vegard,Denton}) we can change the box operator to the gauge invariant form, as discussed in section \ref{intro}
\begin{equation} \label{alloy::eq9}
\Box_j \rightarrow \Box_j + i\alpha c,
\end{equation}
where $\alpha$ is proportional to the expansion coefficient.
Assuming a long-wave modulation of the concentration, we can ignore the higher order correction term in the box operator.

For simplicity, consider a deformed solid with
\begin{equation} \label{alloy::eq10}
\uj = u_s \exp(-i\kvj\cdot \vec{u}(\vec{r})).
\end{equation}
Then we already know that all the local terms in the free energy functional remain invariant under elastic deformations, thus it is sufficient to look at the gradient term.
Thus we get for each wave vector
\begin{equation} \label{alloy::eq11}
[\khj\cdot \nabla + i\alpha c] \uj =-i \uj (\Knorm \khj_m  \khj_l \partial_l u_m - \alpha c)
\end{equation}
Thus the expression in the free energy functional, $|(\khj\cdot \nabla + i\alpha c) \uj|^2$, is minimized for $\alpha c \Knorm^{-1} = \khj_m \khj_l \partial_l u_m$.
Symmetrization gives immediately $\alpha c \Knorm^{-1} = \khj_m \khj_l \epsilon_{lm}$.
Using the orthogonality theorem (\ref{elast::eq4}) we therefore get the relative lattice expansion
\begin{equation} \label{alloy::eq12}
\epsilon_{ll} = d\, \alpha\, c\, \Knorm^{-1}.
\end{equation}
One can readily check (e.g. for the hexagonal lattice (\ref{hex::eq5})) that all diagonal elements of the strain tensor are equal and that the off-diagonal elements vanish, thus we get a dilatational stress free eigenstrain
\begin{equation} \label{alloy::eq13}
\epsilon_{ij}^{0} = \alpha c(\vec{r}) \Knorm^{-1} \delta_{ij}
\end{equation}
and the elastic energy density becomes
\begin{equation} \label{alloy::eq14}
f_{el} = \frac{1}{2} c_{ijkl} (\epsilon_{ij} - \epsilon_{ij}^0) (\epsilon_{kl} - \epsilon_{kl}^0)
\end{equation}
with the same elastic constants as before.

Notice that the lattice dilatation via the modification of the box operator Eq.~(\ref{alloy::eq9}) also affects the impurity diffusion, since it introduces another  concentration dependent term in the free energy.
The ``kinetic'' part of the free energy
\begin{equation} \label{alloy::eq15}
\Delta F_k =-\frac{n_0k_BT C''(\Knorm)}{2}\sum_{j=1}^{N/2} \int d\rv \left| \left(\Box_j +i\alpha c\right) \uj \right|^2
\end{equation}
gives a contribution to the chemical potential,
\begin{eqnarray} 
\mu_k &=& \frac{\delta F_k}{\delta c} = -\frac{n_0k_BT C''(\Knorm)}{2} \times \nonumber \\
&\times& \sum_{j=1}^{N/2} \big[ -2\alpha \khj\cdot \Im(\uj\ujs) -\alpha \Knorm^{-1} \Re(\uj\nabla^2 \ujs) \nonumber \\
&& + 2\alpha^2 c \uj \ujs\big]. \label{alloy::eq16}
\end{eqnarray}
For a deformed solid, Eq.~(\ref{alloy::eq10}), we obtain
\begin{eqnarray}
\mu_k &=& -\frac{n_0k_BT C''(\Knorm)}{2} \big[ -\alpha u_s^2 \Knorm \frac{N}{d} \partial_l u_l \nonumber \\
&& + \frac{1}{2} \alpha u_s^2 \Knorm \frac{N}{d} (\partial_i u_l)(\partial_i u_l) + N \alpha^2 c u_s^2 \big] \label{alloy::eq17}
\end{eqnarray}
where we used the orthogonality relation (\ref{elast::eq4}).
As before, the second term stems from the higher order term in the box operator, and as long as the distortions are small, $\partial_i u_l \ll 1$, it can be neglected (in fact, it corresponds to the nonlinear contribution in the full strain tensor, $\epsilon_{ik}=(\partial_k u_i + \partial_i u_k + \partial_i u_l \partial_k u_l)/2$, which is relevant for its rotational invariance).
This is typically the case if the eigenstrain is small, $\alpha c \Knorm^{-1} \ll 1$.
For a deviation from the equilibrium strain, $\epsilon_{ij} = \epsilon_{ij}^0 + \delta \epsilon_{ij}$, we therefore obtain
\begin{eqnarray}
\mu_k &=& \frac{n_0k_BT C''(\Knorm)}{2} \alpha u_s^2 \Knorm \frac{N}{d} \delta \epsilon_{ll} \nonumber \\
&=& -\Knorm^{-1} \alpha c_{ijkk} (\epsilon_{ij}-\epsilon_{ij}^0) \nonumber \\
&=& -\Knorm^{-1} \alpha \sigma_{kk} \label{alloy::eq20}
\end{eqnarray}
where we used the general expressions (\ref{elast::eq12}) and (\ref{elast::eq13}).
%In particular, for bcc we obtain
%\begin{equation}
%\mu_k = -2\alpha \Knorm^{-1} C_{11} \delta\epsilon_{ll}.
%\end{equation}
Notice that the same expression can also be derived directly from Eq.~(\ref{alloy::eq14}) and $\mu_k = \partial f_{el}/\partial c$.

If the material is stretched, $\delta\epsilon_{ll} > 0$ or $\sigma_{kk}>0$, the chemical potential is reduced, $\mu_k<0$ for a positive Vegard coefficient $\alpha>0$ (the material extends if it contains impurities).
Thus a flow $\vec j \sim -c\nabla\mu_k$ sets in to locally increase the concentration.
Hence, the impurity concentration is affected by the local volume change.
The diffusion equation becomes therefore instead of Eq.~(\ref{alloy::eq5})
\begin{equation} \label{alloy::eq21}
\frac{\partial c}{\partial t} = \nabla\cdot \left[ D \nabla c - D b c \nabla \phi + D \frac{v_0}{R T_M} c \nabla \mu_k \right],
\end{equation}
with the expression for $\mu_k$ being given by Eq.~(\ref{alloy::eq16}).

It is straightforward to generalize the model to cases with more complicated phase diagrams, eventually also using thermodynamic databases.
Similarly, the extension towards nonlinear or nonisotropic lattice expansions and the inclusion of thermal expansion using a temperature dependent expansion coefficient is straightforward.

\subsection{Open-system elastic constants}

The elastic constants $c_{ijkl}$ describe the material stiffness for fixed composition.
As mentioned above, stretching of the material leads to an increased solute concentration and induces elastic relaxation.
Therefore, for fast diffusing elements and slow deformations, the material seems to be effectively softer, because it compensates the elastic deformation by e.g.~filling the interstitial positions with impurities (increase of the stress-free strain $\epsilon_{ij}^0$).
Formally, this leads for fixed chemical potential to the definition of the open system elastic constants\cite{CahnLarche}, which can be calculated readily using the expressions given above.

The chemical potential becomes in the solid with the small strain approximation, see Eqs.~(\ref{alloy::eq1}) and (\ref{alloy::eq20}),
\begin{equation} \label{os::eq1}
\mu = \frac{\delta F}{\delta c} = \frac{R T_M}{v_0} \ln c + \Delta\epsilon -\Knorm^{-1} \alpha \sigma_{kk},
\end{equation}
and therefore the local concentration as function of stress
\begin{equation} \label{os::eq2}
c = c_0 \exp \left[ \frac{v_0 \Knorm^{-1}}{R T_M} \alpha \sigma_{kk} \right]\approx c_0 \left(1 + \frac{v_0 \Knorm^{-1}}{R T_M} \alpha \sigma_{kk} \right)
\end{equation}
with
\begin{equation} \label{os::eq3}
c_0 = \exp \left[ \frac{v_0\Knorm^{-1}}{R T_M} (\mu - \Delta\epsilon) \right].
\end{equation}
For positive Vegard coefficient, $\alpha>0$, Eq.~(\ref{os::eq2}) expresses the local concentration increase under tension, $\sigma_{kk}>0$.
Therefore, the stress-strain relation $\sigma_{ij} = c_{ijkl} (\epsilon_{kl}-\epsilon^0_{kl})$ becomes implicit through the stress free strain (\ref{alloy::eq13})
\begin{equation} \label{os::eq4}
\sigma_{ij} = c_{ijkl} \left[ \epsilon_{kl} - \alpha c_0 \Knorm^{-1} \delta_{kl} - \frac{v_0 c_0}{R T_M \Knorm^{2}} \alpha^2 \delta_{kl} \sigma_{mm}\right].
\end{equation}
This relation can be inverted, 
\begin{equation} \label{os::eq5}
\sigma_{ij} = c_{ijkl}^* [\epsilon_{kl}-\epsilon_{kl}^0(c_0)],
\end{equation}
which defines the  open-system elastic constants $c_{ijkl}^*$.

For a general case of cubic symmetry, with only $c_{xxxx}, c_{xyxy}$ and $c_{xxyy}$ being independent elastic constants, we obtain from Eqs.~(\ref{os::eq4}) and (\ref{os::eq5})
\begin{eqnarray}
c_{xxxx}^* &=& \frac{c_{xxxx} + 2\chi \eta^2 (c_{xxxx} + 2c_{xxyy})(c_{xxxx}-c_{xxyy})}{1+3\chi\eta^2(c_{xxxx}+2c_{xxyy})} \nonumber \\
c_{xxyy}^* &=& \frac{c_{xxyy} + \chi \eta^2 (c_{xxxx} + 2c_{xxyy})(c_{xxyy}-c_{xxxx})}{1+3\chi\eta^2(c_{xxxx}+2c_{xxyy})} \nonumber \\
c_{xyxy}^* &=& c_{xyxy} \label{os::eq6}
\end{eqnarray}
with $\chi$ and $\eta$ being defined as in Ref.~\onlinecite{CahnLarche}
\begin{eqnarray}
\chi &=& \frac{c_0 v_0}{R T_M},\\
\eta &=& \alpha \Knorm^{-1}.
\end{eqnarray}
Interestingly, the combination $I := c_{xxxx}-c_{xxyy}-2c_{xyxy}$ remains the same for the open system, i.e.~$I^*=I$.
Notice that $I=0$ is the condition for isotropy of a cubic system \cite{Landau7}, and since the impurities change the lattice constant uniformly in all directions, the solute does not destroy the isotropy.
In that case, the material can be characterized through two elastic constants, e.g.~the shear modulus $\mu$ and the bulk modulus $K$.
From the general expressions (\ref{os::eq6}) we obtain for $I=0$ the isotropic open system constants
\begin{eqnarray}
\mu^* &=& \mu, \\
\frac{1}{K^*} &=& \frac{1}{K} + 9 \chi \eta^2.
\end{eqnarray}
These expressions match exactly the prediction in Ref.~\onlinecite{CahnLarche}.
The first equation reflects that a pure shear does not change the volume, and therefore no concentration change occurs.
As expected, the open system bulk modulus is smaller than for fixed concentration.
We note that this reduction occurs also for $\alpha<0$, since the correction is quadratic in $\alpha$.

%%%%%%%%%%%%%%%%%%%%%%%%%%%%%%%%%%%%%%%%%%
\section{Validations}

%%%%%%%%%%%%%%%%%%%%%%%%%%%%%%%%%%%%%%%%%%
\subsection{The Asaro-Tiller-Grinfeld instability}
\label{Grinfeld}

\begin{figure}
\begin{center}
\includegraphics[width=6cm]{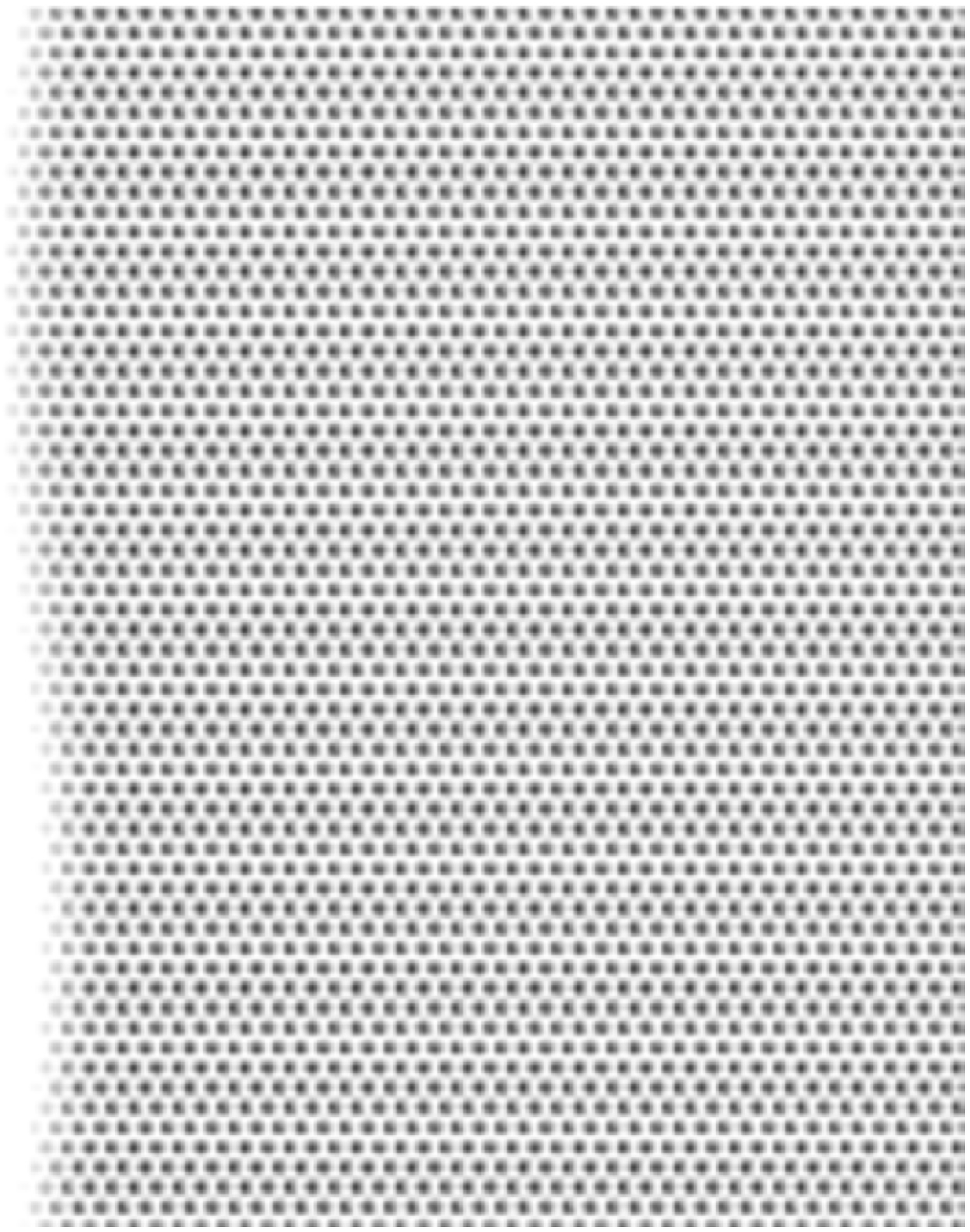} \\[0.5ex]
\includegraphics[width=6cm]{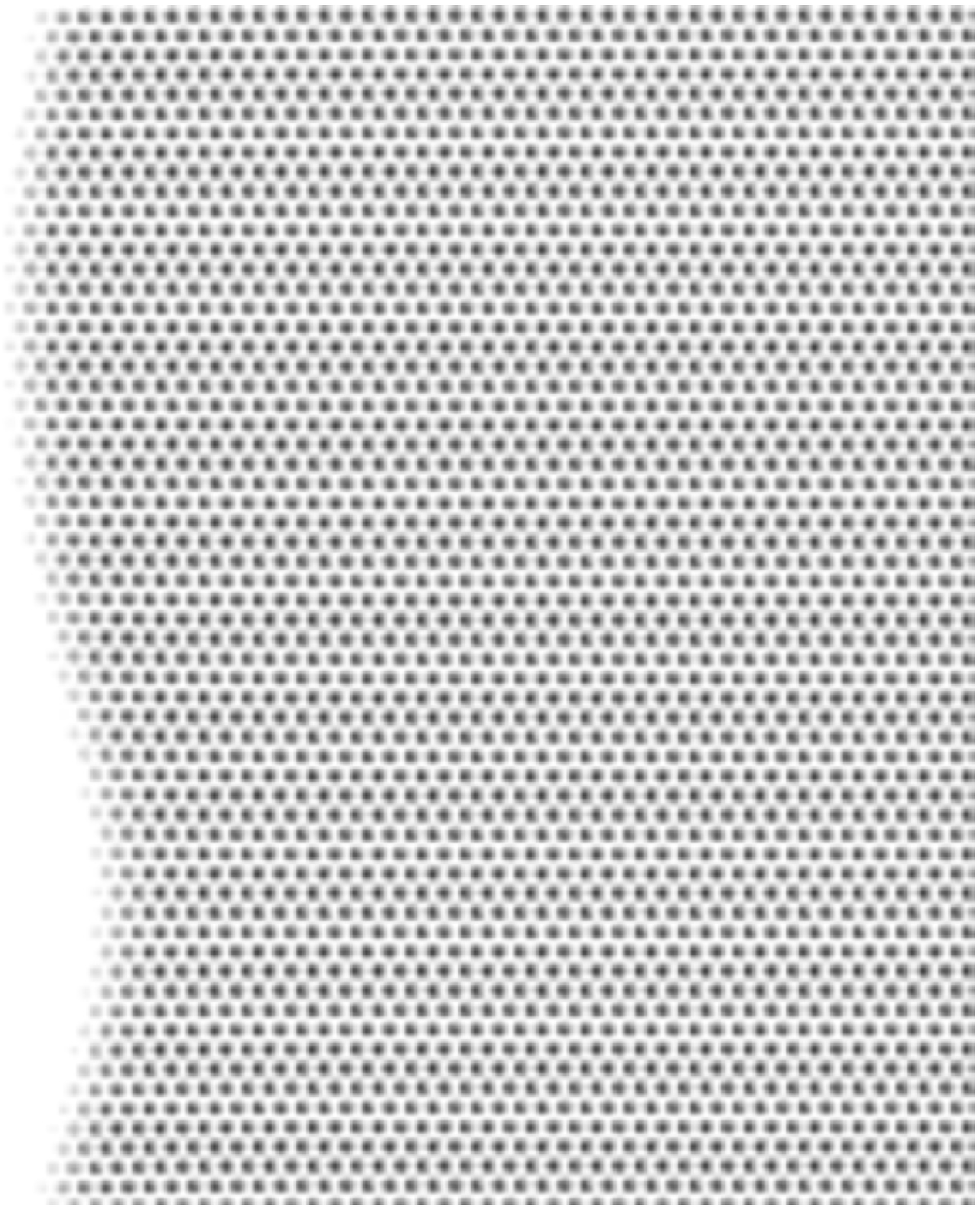} \\[0.5ex]
\includegraphics[width=6cm]{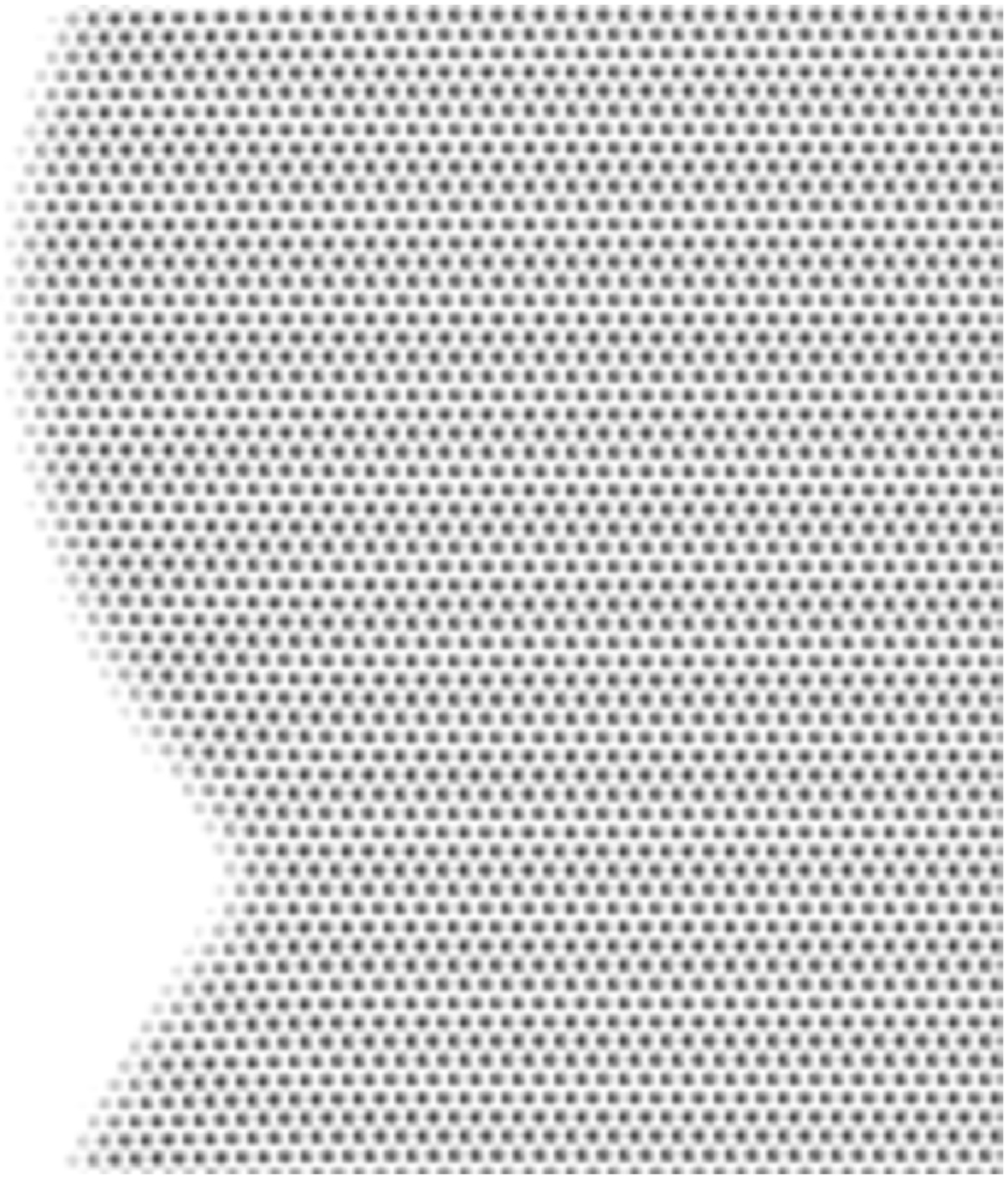} \\[0.5ex]
\caption{Interface evolution of the ATG instability for $kL_G=0.84$ with $K_l/K_s=1$ and $K_s\pfcpropfac=1$.
The ``atomic spacing'' is determined through the parameter $\epspfc$ and is relevant only for the reconstruction of the density waves but does not influence the ``macroscopic'' evolution of the interface. Here
$\epspfc=0.5$ and periodic boundary conditions are used in the vertical direction.
The time in the snapshots is from top to bottom 0, 12000, and 20000, respectively.
At a nonlinear stage of the instability, deep grooves form in the solid to reduce the elastic energy, which can lead to fracture \cite{spatschek2, spatschek1, spatschek3, spatschek4}.
}
\label{Grinfeld::fig2}
\end{center}
\end{figure}

The Asaro-Tiller-Grinfeld (ATG) instability is a morphological instability of a uniaxially strained surface\cite{Asaro,Grinfeld,YanSro93,KasMis94,SpeMer94,Nozieres,Fleck,Wu09}.
The development of corrugations due to a reshuffling of material reduces the total energy for long-wave perturbations.
%The dynamics of this process for a stressed solid phase in contact with its melt was discussed first by Asaro and Tiller.
Here, local melting and solidification at the interface leads to the development of the instability.
In two dimensions, the shape of the interface is described by the profile (see Fig.~\ref{Grinfeld::fig2})
\begin{equation} \label{Grinfeld::eq1}
x(y) = \Delta \sin ky
\end{equation}
and subjected to a tensile or compressive stress $\sigma_0$ along the interface.
Since we assume that the melt phase is stress free, the normal and shear stresses vanish in the solid at the interface.
Then the chemical potential difference at the interface between the solid and the melt becomes in a sharp interface picture \cite{Nozieres}
\begin{equation} \label{Grinfeld::eq4}
\Delta \mu = \Omega (\frac{1}{2} \sigma_{ij}\epsilon_{ij} - \gamma_{\mathrm{eff}}\kappa)
\end{equation}
with the atomic volume $\Omega$ and the interface curvature $\kappa$.
Since the the solid-melt interfacial energy $\gamma(\theta)$ is anisotropic, the stiffness $\gamma_{\mathrm{eff}} = \gamma(\theta) + \gamma''(\theta)$ appears here.

It triggers interface evolution via a melting-solidification process, and the interface normal velocity is given by
\begin{equation} \label{Grinfeld::eq6}
v_n = \frac{\intmob}{\gamma\Omega} \Delta\mu
\end{equation}
with the kinetic coefficient $\intmob$ of the interface kinetics.

The evolution of the interface leads to a time-dependend amplitude (in the framework of a linear stability analysis) $\Delta=\Delta_0\exp\omega t$.
A sharp interface calculation predicts for isotropic elasticity in a two-dimensional plane-strain situation the spectrum 
\begin{equation} \label{Grinfeld::eq2}
\omega = \intmob \left[ 2\frac{1-\nu^2}{E} \sigma_0^2 k - k^2 \gamma_{\mathrm{eff}}(\theta=0) \right],
\end{equation}
where $\intmob$ is the kinetic coefficient of the melting and solidification process, and the (planar) interface normal direction is assumed to correspond to $\theta=0$.
$E$ and $\nu$ are Young's modulus and Poisson ratio respectively.
Here, the first term accounts for the elastic destabilization, whereas the second term describes the stabilization due surface energy.
%Although we assume that the elastic problem is isotropic (as it is the case for the hexagonal system), the surface energy is usually still anisotropic.
The above spectrum sets a characteristic lengthscale, the Grinfeld length,
\begin{equation} \label{Grinfeld::eq5}
L_G = \frac{E\gamma_{\mathrm{eff}}}{2(1-\nu^2)\sigma_0^2}.
\end{equation}

We model this process using the two-dimensional amplitude equations for a hexagonal system and also use this notation (see Appendix \ref{hex}).
We note that the sixfold symmetry induces elastic isotropy, whereas the interfacial energy remains anisotropic (see also Fig.~\ref{box::fig2} in Section \ref{box}).
For the amplitude equations we use relaxation equations of the type
\begin{equation} \label{Grinfeld::eq3}
\frac{\partial A_j^0}{\partial t} = - K_j \frac{\delta F}{\delta A_j^{0*}},
\end{equation}
with kinetic coefficients $K_j$.
These equations do not lead to the same sharp interface limit as used for the ATG spectrum above if all $K_j=K$ are constants.
The reason is that the motion of the interface occurs on the same timescale as the relaxation of the elastic degrees of freedom.
This is problematic especially in the long wave limit, because the range of the elastic distortion is the same as the wavelength of the interface corrugation, and the elastic fields have to adjust via a diffusive process.
In reality, however, the interface motion is slow in comparison to the sound speed (which sets the true scale for the elastic relaxation), and therefore the assumption of static elasticity as in Eq.~(\ref{Grinfeld::eq2}) is appropriate.
%We note that the behavior differs from phase field crystal simulations, where the interface motion is slow due to mass conservation, thus static elasticity can be established sufficiently fast;
%the interface evolution, however, is limited by bulk transport \cite{Wu09}.

To obtain the same behavior with the amplitude equations, we use a kinetic coefficient that depends on the amplitudes:
In the solid, the kinetic coefficient $K_j=K_s$ is high and low in the liquid, $K_j=K_l$.
In between, the coefficients are interpolated,
\begin{equation}
K_j = h_j K_s + (1-h_j) K_l,
\end{equation}
with interpolation functions $h_j$ that have value $1$ in the solid and $0$ in the liquid.
This is the aforementioned dependence of the kinetic coefficient $\Gamma$ (in the notation of Section \ref{intro}) on the local values of the amplitudes.
For sharp interfaces, i.e. the width of the diffuse interfaces $\sim \epspfc^{-1/2}$ being small in comparison to the wavelength of the perturbation $\sim 1/k$, the precise choice of the interpolation is not crucial.
In particular, we used $h_j=h(|A_j^0/A_s^0|^2)$, with $h(x)$ being given by Eq.~(\ref{dw::eq5c}).
%\comment{So for the simulations I had indeed used the ``quartic'' interpolation.}
Since the motion of the interface is basically determined by the smaller of the coefficients $K_s, K_l$, we can therefore get a slow motion of the interface, whereas the elastic relaxation in the solid is sufficiently fast, since it is determined by $K_s$.
In the limit $K_l/K_s\to 0$ we therefore recover the case of quasistatic elasticity.

Since the kinetic coefficient $\intmob$ cannot easily be expressed in terms of the mobilities $K_s$ and $K_l$, we first investigated the decay of capillary waves without elastic effects, i.e.~without application of an external stress.
From the decay rate and the spectrum (\ref{Grinfeld::eq2}) we therefore extract the value of $\intmob$ for given values of $K_s, K_l$.
Next, we apply additionally an elastic deformation tangentially to the interface and measure the modified amplitude evolution.
Snapshots of the temporal evolution in the unstable regime are shown in \ref{Grinfeld::fig2}.
The simulations are started with a small initial amplitude $k\Delta_0\approx 0.11$, which grows here for $kL_G=0.84$.
The interface thickness $\xi$ used in the simulations is related to the wavelength of the perturbation by $k\xi \approx 0.3$.

We use a straightforward real space discretization of the amplitude equations, and fixed boundary conditions in the direction perpendicular to the interface ($x$ direction, see Fig.~\ref{Grinfeld::fig2}):
In the liquid ($x=0$) the amplitudes are fixed to zero, whereas at the right interface ($x=X$) we have
\begin{eqnarray*}
A_1(X, y) &=& A_s \exp[-i\khn{1}\cdot \vec{u}_0(X, y)], \\
A_2(X, y) &=& -A_s \exp[-i\khn{2}\cdot \vec{u}_0(X, y)], \\
A_3(X, y) &=& A_s \exp[-i\khn{3}\cdot \vec{u}_0(X, y)],
\end{eqnarray*}
with the homogeneous displacement field for the planar front
\[
\vec{u}_0(x, y) = \left(
\begin{array}{c}
x \epsilon_{xx}^0\\
y \epsilon_{yy}^0
\end{array}
\right),
\] 
which depends on the homogeneous strains $\epsilon_{xx}^0$ and $\epsilon_{yy}^0$.
Notice that for a homogeneous nonhydrostatic stress $\sigma_0$ the system needs to be strained in both directions according to
\[
\epsilon_{xx}^0 = -\frac{\nu(1+\nu)}{E} \sigma_0,\qquad \epsilon_{yy}^0=\frac{1-\nu^2}{E} \sigma_0.
\]

In the other direction along the interface ($y$ direction), we use quasiperiodic boundary conditions:
\[
A_j(x, Y) = A_j(x, 0) \exp[-i\khn{j}\cdot\Delta \vec{u}_0]
\]
where $Y$ is the system size in this direction, and the displacement jump $\Delta\vec{u}_0 = \vec{u}_0(x, Y)-\vec{u}_0(x, 0)$.
The wavelength of the perturbation therefore has to fit into this periodic interval.
Instead of changing the wavelength to scan the spectrum of the ATG instability, we vary the stress $\sigma_0$;
this also has the advantage that the scale separation between the different geometrical length scales does not change.

Due to the energy increase of the solid through the mechanical load, it is no longer in equilibrium with the melt at $T=T_M$.
It is therefore convenient to suppress the planar front motion by a slight undercooling, since we focus here on the development of the instability.

\begin{figure}
\begin{center}
\includegraphics[width=8cm]{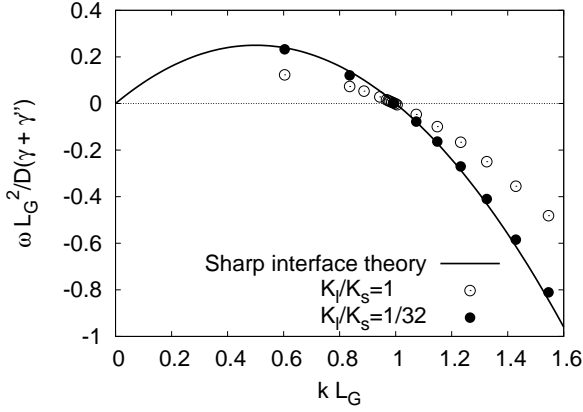}
\caption{Spectrum of the ATG instability. The solid line is the analytical sharp interface prediction Eq.~(\ref{Grinfeld::eq2}), the dots show the numerical results obtained from the amplitude equations, for two different mobility ratios. For small $K_l/K_s$ we recover the sharp interface limit because the strain relaxes fast in comparison to the interface motion.}
\label{Grinfeld::fig1}
\end{center}
\end{figure}

For the calculation of the spectrum, we use only the first linear regime of the amplitude evolution (after an initial stage where the interfaces adjust to the proper profiles).
The results are shown in Fig.~\ref{Grinfeld::fig1}, together with a comparison to the sharp interface prediction (\ref{Grinfeld::eq2}).
We clearly see that in the limit $K_l/K_s\to 0$ the spectrum agrees well with the analytical theory.
In particular, for $kL_G\approx 1$ we get $\omega\approx0$;
then the interface motion is slow, and therefore even for equal mobility in solid and melt the elastic fields can adjust fast enough.
Hence curves for different mobility ratios intersect all at $kL_G=1$.

Since the rotation of the grains is not important here, the higher order correction in the box operator can be neglected.
We checked numerically that for $\epspfc=0.1$ the correction term gives only negligible modifications of the results.
Then the small parameter $\epspfc$ appears in the free energy functional Eq.~(\ref{hex::eq22}) and in the amplitude equations (\ref{Grinfeld::eq3}) only as multiplicative constant and can be absorbed in the kinetic coefficients, thus the description is entirely on the slow scale $\Rv$.
The parameter $\epspfc$ comes in only via the reconstruction of the density waves according to Eq.~(\ref{hex::eq18}).
In particular, for a solid that is not rotated the amplitudes are constant in the solid (without strain) or vary only gently if a strain is present, and therefore the amplitudes can be discretized on a scale that is independent of the ``atomic'' resolution.
In this sense, the computational efficiency of the model is not inferior to a conventional phase field model, apart from the fact that more than one parameter is needed.
On the other hand, the description automatically contains elasticity, which would require a separate treatment in a conventional model, see e.g.~Ref.~\onlinecite{spatschek3}.

%%%%%%%%%%%%%%%%%%%%%%%%%%%%%%%%%%%%%%%%%%
%%%%%%%%%%%%%%%%%%%%%%%%%%%%%%%%%%%%%%%%%%
\subsection{Crystal-melt interfacial free energies and polycrystalline growth}
\label{box}

A central part of the theory is the box operator, which generalizes the gradient term $\khj\cdot \nabla$ of a more conventional Ginzburg-Landau theory to a rotational invariant form,
\begin{equation} \label{box::eq1}
\Boxnew_j = \khj\cdot\nablanew - \frac{i\epsnew^{1/2}}{2} \nablanew^2,
\end{equation}
which introduces a higher order correction.
We use here the dimensionless representation introduced in Eqs.~(\ref{dw::eq6})-(\ref{dw::eq12}).
Since it also brings higher order derivatives, a numerical treatment becomes computationally more costly in an explicit scheme, since then the relaxation timesteps have to be rather short.

\begin{figure}
\begin{center}
\includegraphics[width=8cm]{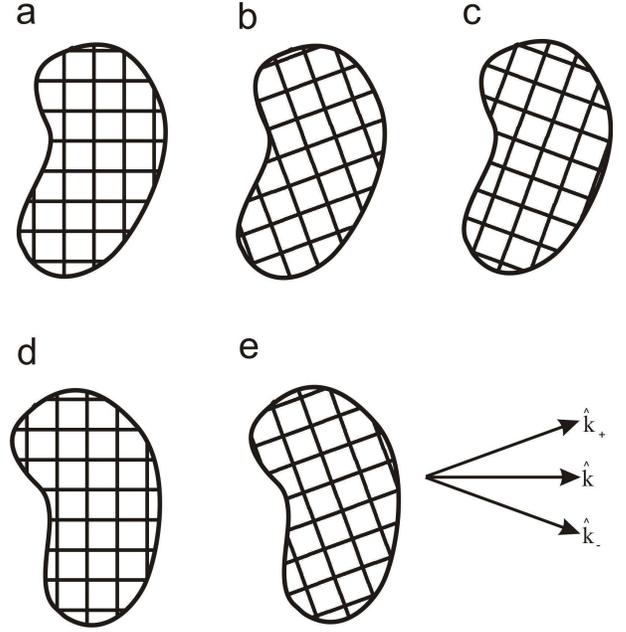}
\caption{Schematic illustration of different rotations of a crystal and/or its interface contour. (a) Original state with amplitude $\Anew(\Rnew)$. (b) The contour of the crystal is the same, but the lattice is rotated; the amplitude is $\Anew(\Rnew)\exp(i\kh^\dagger \mathbf{M} \Rnew/\epsnew^{1/2})$; (c) The contour is the same, but the lattice  is rotated in the opposite direction; $\Anew(\Rnew)\exp(i\kh^\dagger \mathbf{M}^\dagger \Rnew/\epsnew^{1/2})$. (d) The contour is rotated, but the lattice orientation is the same as in (a); the amplitude is $\Anew(\mathbf{R}\Rnew)$. (e) Both the contour and the lattice are rotated; therefore this state is equivalent to (a) and the amplitude is $\Anew(\mathbf{R}\Rnew)\exp(i\kh^\dagger \mathbf{M} \Rnew/\epsnew^{1/2})$.
}
\label{box::fig1}
\end{center}
\end{figure}

In the following, we pick a particular k-vector, and drop therefore the subscript $j$.
In a pure solid phase with a spatially constant amplitude $A$, we have $\Boxnew \Anew=0$.
However, this relation also holds if we describe a solid in a rotated state.
Namely, if $\Anew(\Rnew)$ is an amplitude field, a crystal with the same shape, but rotated grain structure, as sketched in the transition from Fig. \ref{box::fig1}a to b, is described by
\begin{equation} \label{box::eq2}
\rotp{\Anew}(\Rnew) = \Anew(\Rnew) \exp\frac{i\kh^\dagger \mathbf{M} \Rnew}{\epsnew^{1/2}},
\end{equation}
where the dagger $\dagger$ denotes transposition.
Here, $\mathbf{M}=\mathbf{R}-\mathbf{I}$, where $\mathbf{I}$ is the unity matrix and $\mathbf{R}$ an orthogonal rotation matrix.
In two dimensions, $\mathbf{R}$ has therefore the structure
\begin{equation} \label{box::eq3}
\mathbf{R} = \left(
\begin{array}{cc}
\cos\theta & \sin\theta \\
-\sin\theta & \cos\theta
\end{array}
\right).
\end{equation}
The reason is that the rotated lattice structure is described by the density field
\begin{equation}
\rotp{n}(\Rnew) = \Anew(\Rnew) \exp\frac{i\kh^\dagger \mathbf{R} \Rnew}{\epsnew^{1/2}} = \Anew(\Rnew) \exp\frac{i\kh^\dagger \mathbf{M} \Rnew}{\epsnew^{1/2}} \exp\frac{i\kh^\dagger \Rnew}{\epsnew^{1/2}},
\end{equation}
where is the first step the rotated k-vector is $\rotp{\kh} = \mathbf{R}^\dagger \kh$, and in the second step we separated the fast oscillating factor $\exp(i\kh^\dagger \Rnew/\epsnew^{1/2})$ in the spirit of the multiscale expansion.
The first two factors are therefore the amplitude with respect to the basis set of the original k-vectors.
 
The first important property of the box operator is \cite{Gunetal94,Gra96}
\begin{equation} \label{boxtheorem1}
\Boxnew \exp\frac{i\kh^\dagger  \mathbf{M}\Rnew}{\epsnew^{1/2}} = 0.
\end{equation}
It implies that a pure crystal is a solution of the amplitude equations for arbitrary orientation.
We note that we use both the notation of a scalar product (denoted by a dot $\cdot$) and a matrix product (no multiplication symbol), i.e. $\vec{a}\cdot\vec{b} = \vec{a}^\dagger \vec{b}$.
From the definition of the operator we obtain
\begin{eqnarray*}
\Boxnew \exp\frac{i\kh^\dagger  \mathbf{M}\Rnew}{\epsnew^{1/2}} &=& \epsnew^{-1/2} \left[ i\kh^\dagger \mathbf{M} \kh + \frac{1}{2} i (\mathrm{M}^\dagger \kh)^\dagger \mathbf{M}^\dagger \kh \right] \times \\
&& \times \exp\frac{i\kh^\dagger  \mathbf{M}\Rnew}{\epsnew^{1/2}} \\
&=& \frac{i}{\epsnew^{1/2}} \left[ \kh^\dagger (\mathbf{R}-\mathbf{I}) \kh + \frac{1}{2} \kh^\dagger (\mathbf{R}-\mathbf{I}) (\mathbf{R}^\dagger -\mathbf{I}) \kh \right] \times \\
&& \times \exp\frac{i\kh^\dagger  \mathbf{M}\Rnew}{\epsnew^{1/2}} \\
&=& 0
\end{eqnarray*}
where we used the normalization condition $|\kh|=1$ and the orthogonality of the rotation matrix, $\mathbf{R}^\dagger\mathbf{R}=1$.

Next, we consider situations in which the whole solid is rotated, but the lattice orientation is kept in its original state.
This is visualized in Fig.~\ref{box::fig1}a and d.
We introduce rotated lattice vectors $\rotm{\kh} = \mathbf{R} \kh$;
notice that in comparison to $\rotp{\kh}$ this vector is rotated in the opposite direction.
Correspondingly, we define a rotated box operator
\begin{equation}
\rotm{\Boxnew} = \rotm{\kh}\cdot\nablanew - \frac{i\epsnew^{1/2}}{2} \nablanew^2.
\end{equation}
If $\Anew(\Rnew)$ is a density wave amplitude, then $\Anew(\mathbf{R}\Rnew)$ describes the crystal with rotated shape, but the same lattice orientation.
We obtain then
\begin{equation} \label{boxtheorem2}
\Boxnew \Anew(\mathbf{R}\Rnew) = \left. \rotm{\Boxnew} \Anew\right|_{\mathbf{R}\Rnew},
\end{equation}
which expresses the equivalence of {\em active} and {\em passive} rotations:
The rotated grain with original lattice orientation (Fig.~\ref{box::fig1}d) has the same properties as the original grain with a lattice that is rotated in opposite direction (Fig.~\ref{box::fig1}c).
To obtain the relation (\ref{boxtheorem2}) we first note that $\nablanew \Anew({\mathbf R} \Rnew) = \left. \mathbf{R}^\dagger \nablanew \Anew\right|_{\mathbf{R} \Rnew}$.
Furthermore, the Laplace operator is rotational invariant, i.e. $\nablanew^2 \Anew({\mathbf R} \Rnew) = \left.\nablanew^2 \Anew\right|_{\mathbf{R} \Rnew}$.
Then we get
\begin{eqnarray*}
\Boxnew \Anew({\mathbf R} \Rnew) &=& \kh^\dagger \mathbf{R}^\dagger \left. \nablanew \Anew \right|_{\mathbf{R} \Rnew} - \frac{i\epsnew^{1/2}}{2} \left. \nablanew^2 \Anew\right|_{\mathbf{R} \Rnew} \\
&=& \rotm{\kh} \cdot \left. \nablanew \Anew \right|_{\mathbf{R} \Rnew} - \frac{i\epsnew^{1/2}}{2} \left. \nablanew^2 \Anew\right|_{\mathbf{R} \Rnew} \\
&=& \left. \rotm{\Boxnew} \Anew\right|_{\mathbf{R}\Rnew}.
\end{eqnarray*}

From the definition of the box operator follows immediately the product rule
\begin{equation} \label{boxtheorem3}
\Boxnew(f g) = f\Boxnew g + g\Boxnew f - i\epsnew^{1/2} (\nablanew f)\cdot(\nablanew g).
\end{equation}

We can also define a box operator with opposite rotation
\begin{equation}
\rotp{\Boxnew} = \rotp{\kh}\cdot\nablanew - \frac{i\epsnew^{1/2}}{2} \nablanew^2,
\end{equation}
and from the product rule Eq.~(\ref{boxtheorem3}) and Eq.~(\ref{boxtheorem1}) we obtain the operator rotation rule
\begin{equation} \label{boxtheorem4}
\Boxnew\left( \Anew \exp\frac{i\kh^\dagger \mathbf{M}\Rnew}{\epsnew^{1/2}} \right) =  \exp\frac{i\kh^\dagger \mathbf{M}\Rnew}{\epsnew^{1/2}} \rotp{\Boxnew} \Anew,
\end{equation}
or, with the inverse rotation
\begin{equation} \label{boxtheorem5}
\Boxnew\left( \Anew \exp\frac{i\kh^\dagger \mathbf{M}^\dagger \Rnew}{\epsnew^{1/2}} \right) =  \exp\frac{i\kh^\dagger \mathbf{M}^\dagger\Rnew}{\epsnew^{1/2}} \rotm{\Boxnew} \Anew.
\end{equation}
Analogous to Eq.~(\ref{boxtheorem2}) we have
\begin{equation} \label{boxtheorem6}
\rotp{\Boxnew} \Anew(\mathbf{R}\Rnew) = \left. \Boxnew \Anew\right|_{\mathbf{R}\Rnew}.
\end{equation}
Using Eqs.~(\ref{boxtheorem4}) and (\ref{boxtheorem6}) we finally get the coordinate transformation rules towards a rotated frame of reference, i.e.~the transition from a to e in Fig.~\ref{box::fig1},
\begin{equation}
\Boxnew \left( \Anew(\mathbf{R}\Rnew) \exp\frac{i\kh^\dagger \mathbf{M} \Rnew}{\epsnew^{1/2}} \right) = \exp\frac{i\kh^\dagger \mathbf{M} \Rnew}{\epsnew^{1/2}} \left. \Boxnew \Anew \right|_{\mathbf{R}\Rnew} 
\end{equation}
and
\begin{equation}
\Boxnew^2 \left( \Anew(\mathbf{R}\Rnew) \exp\frac{i\kh^\dagger \mathbf{M} \Rnew}{\epsnew^{1/2}} \right) = \exp\frac{i\kh^\dagger \mathbf{M} \Rnew}{\epsnew^{1/2}} \left. \Boxnew^2 \Anew \right|_{\mathbf{R}\Rnew}.
\end{equation}
The latter equation expresses the rotational invariance of the amplitude equations. Namely,
if $\Anew(\Rnew)$ is a valid amplitude field, the field $\Anew(\mathbf{R} \Rnew)$ describes a rotated material, but still with the same lattice orientation as before.
The multiplication with the exponential factor corresponds to the rotation of the lattice only, and therefore the expression in brackets on the left hand side describes the rotated material.
As expressed by this relation, we obtain the physical equivalence of the state, and therefore the rotational invariance.
Notice that all local terms also acquire the same exponential rotation factor, and therefore this factor cancels in the end in the homogeneous amplitude equations.

We therefore conclude that also the free energy remains unchanged by a rotation of both the lattice and the microstructure.
This, in contrast, is not true without the corrective term of the box operator, since the operator $\kh\cdot\nablanew$ is {\em not} rotational invariant \cite{Gunetal94,Gra96}.
If we consider e.g. a planar solid-melt interface, which is stable exactly at the melting point, independent of the interface normal direction.
The rotational invariant formulation with the box operator preserves this if the whole system is rotated (transition a to d in Fig.~\ref{box::fig1}), where the amplitudes acquire apart from the rotation of the microstructure profile also the beats, i.e. $\Anew(\Rnew)\to \Anew(\mathbf{R}\Rnew) \exp(i\kh^\dagger \mathbf{M} \Rnew/\epsnew^{1/2})$;
the energy of the solid phase remains unchanged (this is trivial for the melt, since there the amplitudes vanish and are therefore always invariant).
Without the higher order term in the box operator, however, the energy density of the solid {\em increases} spuriously, and therefore the solid would start to melt.

\begin{figure}
\begin{center}
\includegraphics[width=8cm]{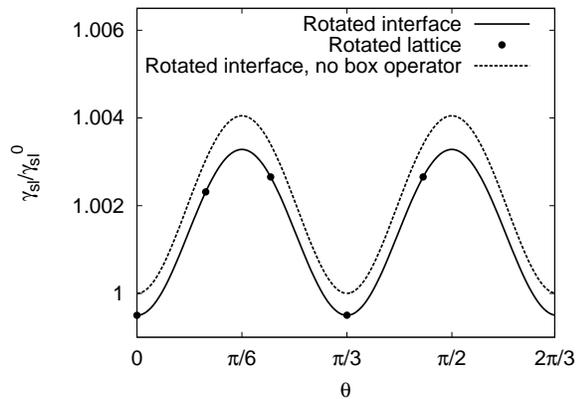}
\caption{Solid-liquid interface energy as function of orientation for the two-dimensional hexagonal system. For $\epspfc=0.1$ the influence of the higher order term in the box operator gives only a small correction. The curves show results from a rotation of the interface normal vector, whereas the points show results for differently rotated lattices. The results coincide if the box operator is included; without it, a rotated lattice structure has a higher bulk energy and is therefore not in equilibrium with the liquid at the nominal melting temperature. $\gamma_{sl}^0$ is the solid-liquid interfacial energy for $\theta=0$ and $\epspfc=0$, i.e.~without the higher order correction of the box operator.}
\label{box::fig2}
\end{center}
\end{figure}

To make this more transparent, we calculated the anisotropic surface energy density for the two-dimensional hexagonal system as function of the interface orientation, as shown in Fig.~\ref{box::fig2}.
The curves are obtained from equilibration runs to minimize the free energy with fixed k-vectors, but with different interface normal vectors, as done in Refs.~\onlinecite{Wu06,Wu07}.
Since we assume a straight interface, all amplitudes depend only on the normal direction, and the problem becomes one-dimensional.
The dashed curve shows the result without correction term in the box operator (i.e. formally setting $\epsnew=0$), the solid curve for finite, small $\epsnew$.
Both curves differ only very little, in agreement with the fact that the higher order term in the box operator gives only a small correction.
Notice that since we do not rotate the reciprocal lattice vectors but only the normal vector, the solid-liquid interface remains stable at $T=T_M$.
The points, in contrast, show data with rotated reciprocal lattice vectors and fixed interface normal.
As expected, the results fall exactly onto the curve with fixed reciprocal lattice vectors and rotated interface normal.
Without the box operator corrections, the equilibration would lead to a pure liquid (no phase coexistence), since the solid bulk energy would be raised artificially, thus making the solid unfavorable at the nominal melting temperature;
this behavior would obviously be unphysical.
We note that the simulations with rotated lattice vectors require the solution of the full two-dimensional problem, since the amplitudes depend now on both coordinates due to the beats of the exponential factor.

Table \ref{box::table1} lists the equilibrium interfacial free energies between solid and melt for bcc iron, using the parameters shown in table \ref{elast::table1}.
Here we clearly see that the higher order term gives only a small correction to the values calculated in Ref.~\onlinecite{Wu07}.
\begin{table}
\caption{Solid-melt interfacial free energy for different interface orientations of bcc iron. The value are given in $\mathrm{J}/\mathrm{m}^2$.}
\label{box::table1}
\begin{tabular}{c|c|c}
\hline\hline
Orientation& without box operator & with box operator \\
\hline
100 & $0.14414$ & $0.14392$ \\
110 & $0.14067$ & $0.14051$ \\
111 & $0.13576$ & $0.13643$ \\
\hline\hline
\end{tabular}
\end{table}
%
%So far we have seen that a crystal can be rotated by an arbitrary angle, which leads to beats in the density wave amplitudes, but still this states solves the equilibrium equations and, correspondingly, the free energy remains unchanged.
%To be precise, we denote this free energy by $F_{AE}[\{A_i\}]$, which is a functional of all amplitudes.
%In contrast $F[n]$ is the original free energy expression either from density functional theory or phase field crystal.

\begin{figure*}
\begin{center}
\includegraphics[width=8cm]{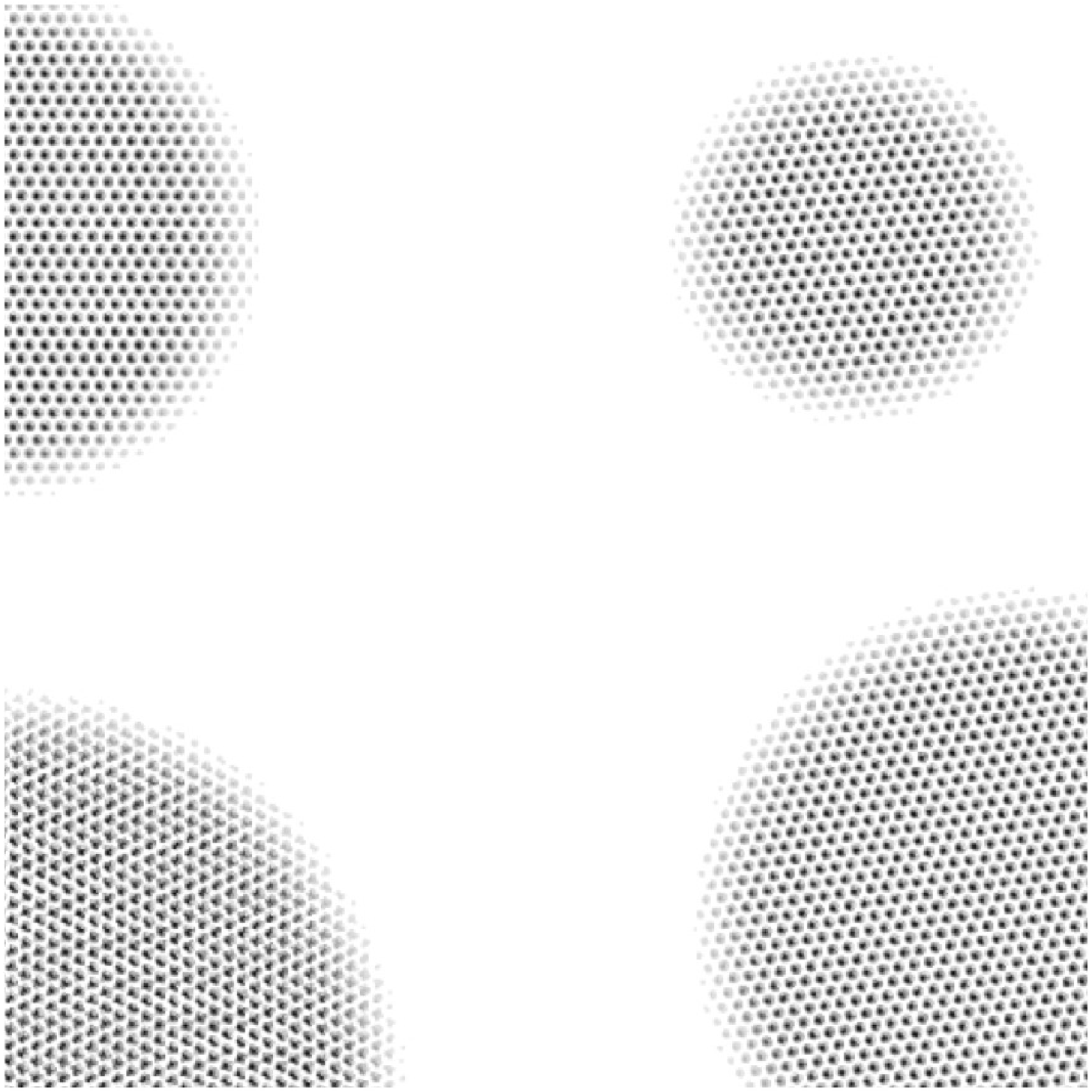}
\includegraphics[width=8cm]{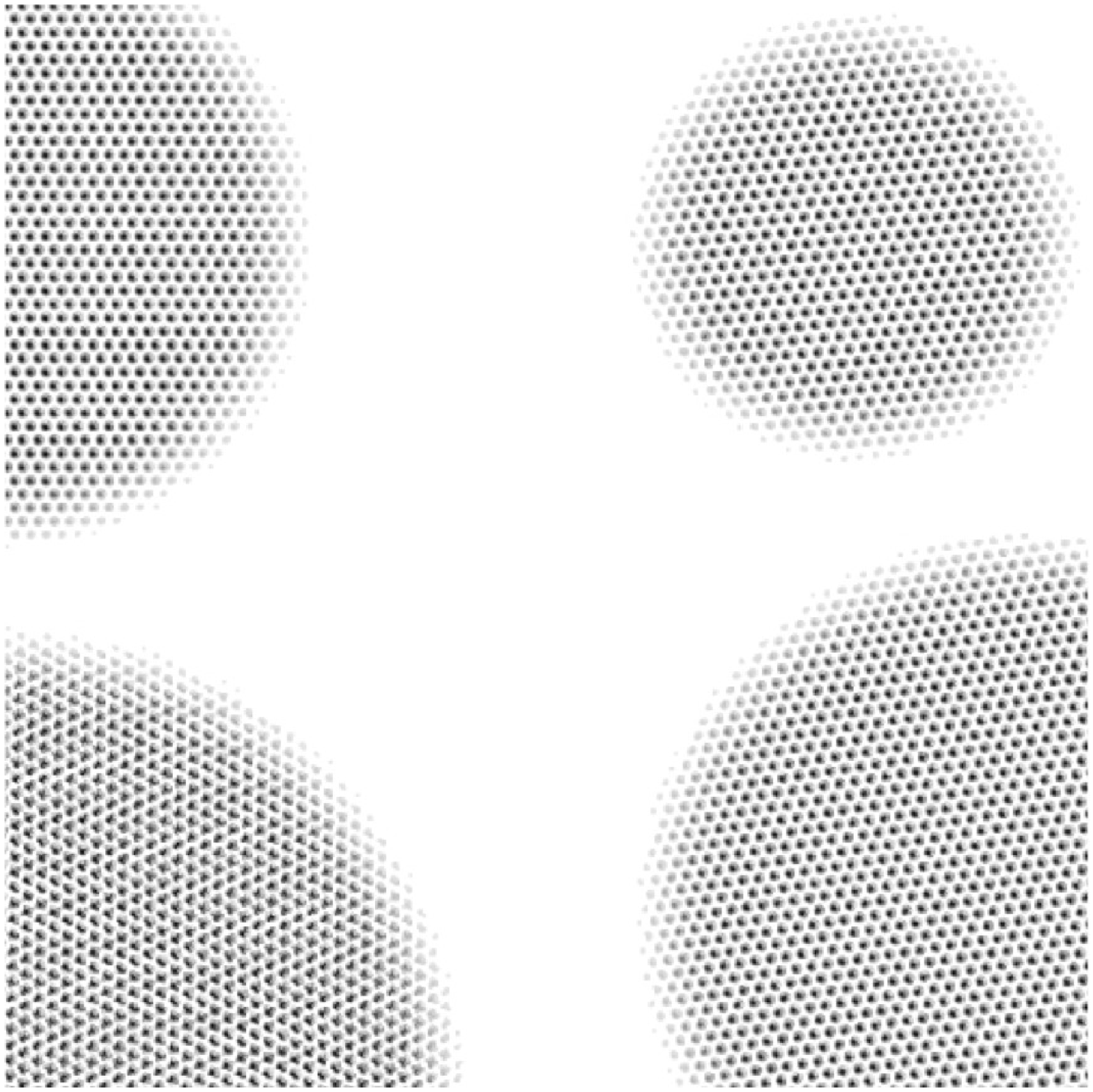}
\includegraphics[width=8cm]{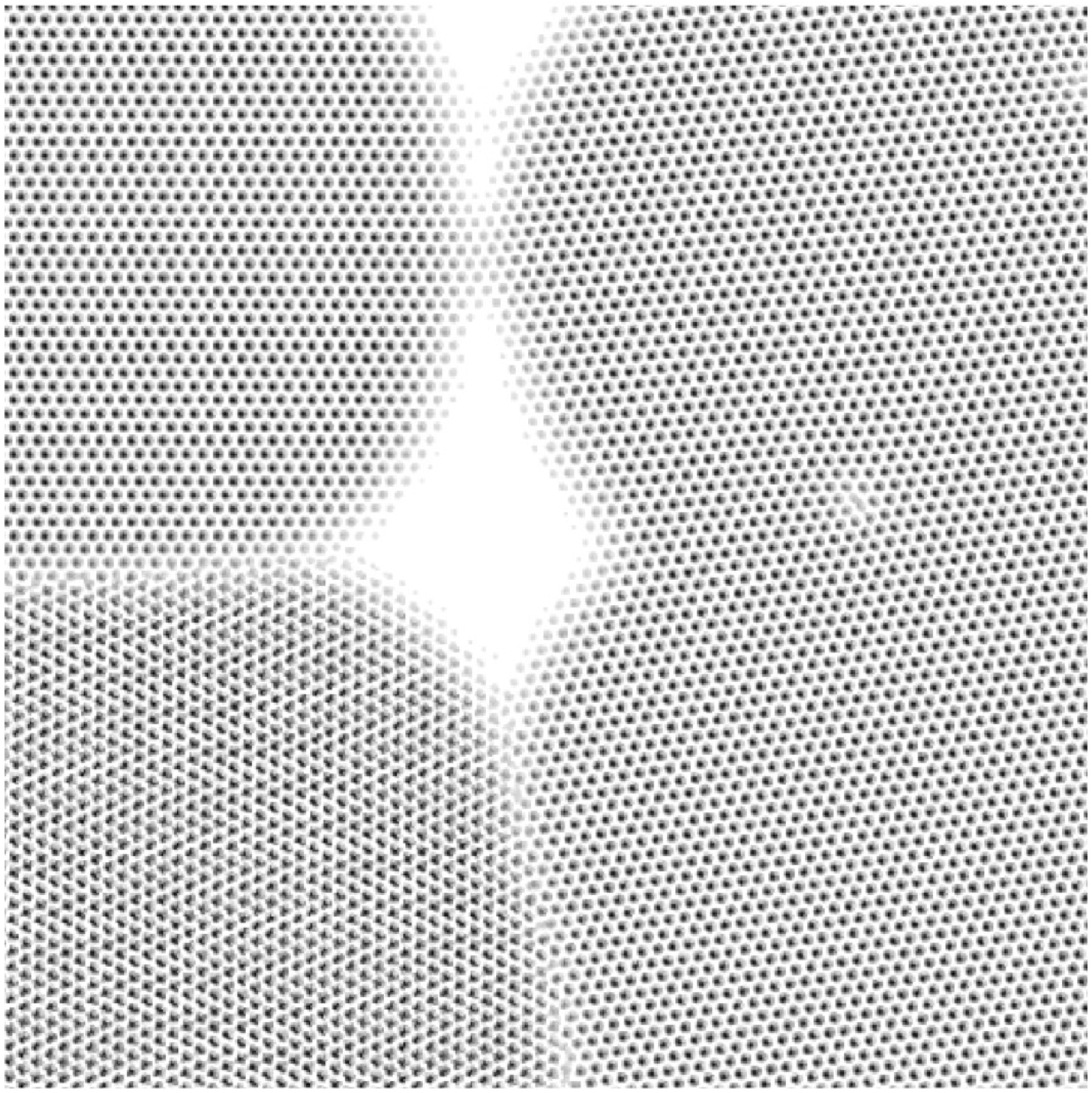}
\includegraphics[width=8cm]{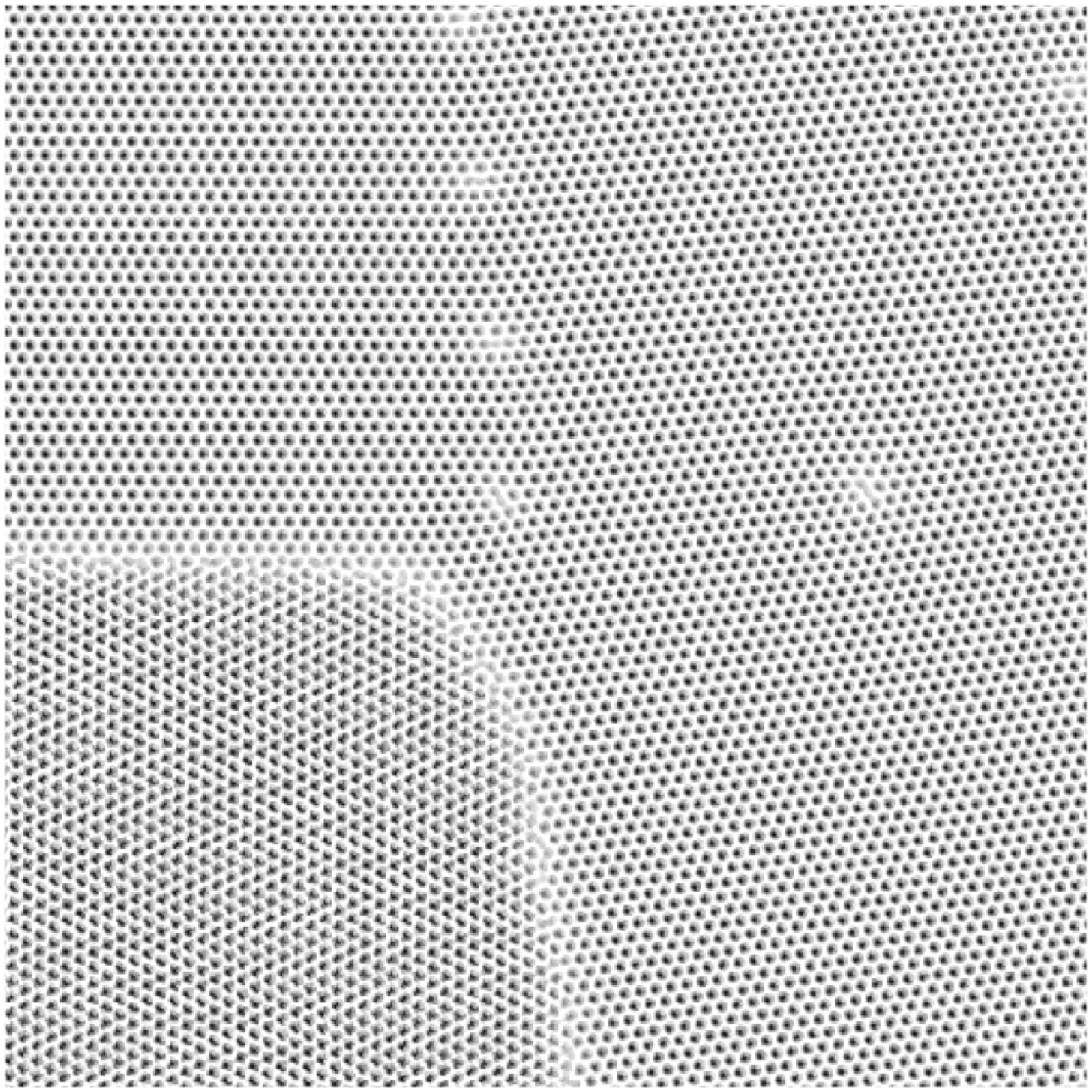}
\caption{Illustration of polycrystalline isothermal solidification of two-dimensional, pure hexagonal crystals, modeled by the amplitude equations of the phase field model, as given in Appendix \ref{hex}. The actual system is much larger than the magnification shown here. When interfaces between misoriented grains form, they generate isolated dislocations for subgrain boundaries and diffuse and premelted interfaces for high misorientations. The parameters are $\Tpfc=-0.002$ and $\epspfc=0.1$.}
\label{poly::fig1}
\end{center}
\end{figure*}

It follows from the frame invariance of the crystal-melt interfacial free-energies, that the amplitude equation approach should describe well the solidification of a polycrystalline material from an undercooled melt. We illustrate this here for the case of two-dimensional hexagonal crystals (see Fig.~\ref{poly::fig1}).
Several spherical seed crystals with different orientation are implanted into the melt phase and grow (provided that they exceed the critical radius).
When the crystals meet, they form grain boundaries, which can consist of isolated dislocations for low angle grain boundaries or show a rather diffuse interface region, which can be partially premelted.
Notice that the defect distribution is not static but slowly evolves, since the dislocations interact with each other via long-range elastic forces.

\subsection{Grain boundary energies and premelting}

Up to this point, the amplitude equations reflect the rotation invariance of the physical system correctly, and this is related to the fact that the melt is fully rotational invariant, since all amplitudes vanish there.
The situation becomes more complex if we consider a polycrystal.
Let us consider a grain boundary, where the energy depends on the orientation of both crystals.
E.g.~for a hexagonal crystal, it is obvious that apart from the {\em continuous} symmetries to which we paid attention so far, also {\em discrete} symmetries are important:
If we rotate one of the adjacent crystals by $60^\circ$, it is in the same state again, and therefore the grain boundary energy has not changed -- it exhibits a sixfold symmetry.

\begin{figure}
\begin{center}
\includegraphics[width=8cm]{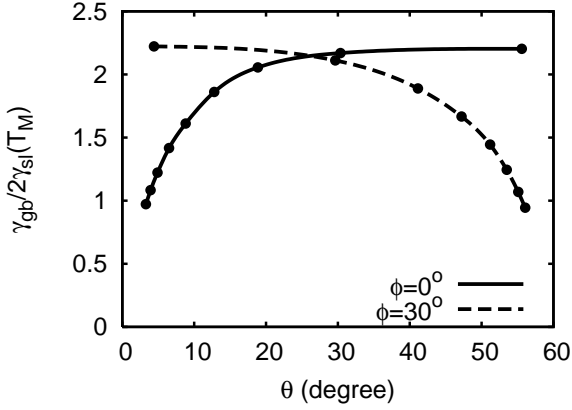}
\caption{Grain boundary energy as a function of misorientation for symmetrical tilt boundaries in a hexagonal crystal  for two different inclinations, normalized to the solid-liquid interfacial energy. The dimensionless undercooling is $\Tpfc=-0.01$, and $\epspfc=0.1$. The results do not obey sixfold symmetry.
%The thin dotted line indicated how the true grain boundary energy may look like.
}
\label{rotlim::fig3}
\end{center}
\end{figure}

The dependence of the grain boundary energy on misorientation 
is shown in Fig.~\ref{rotlim::fig3} for a symmetric tilt boundary in a hexagonal crystal.
The temperature is appreciably below the melting point, in order to ``stabilize'' the grain boundary and to prevent a large separation of the grains due to premelting, which will be briefly discussed below.
\begin{figure}
\begin{center}
\includegraphics[width=8cm]{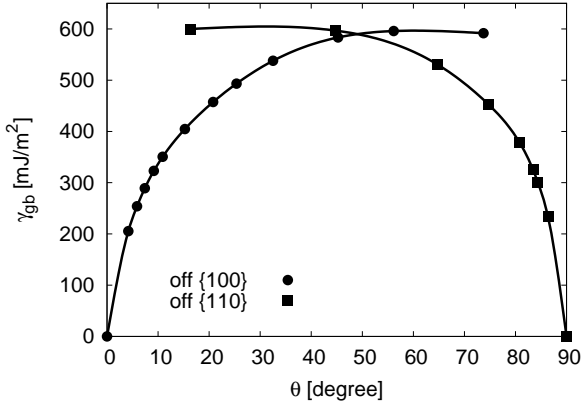}
\caption{Grain boundary energy as function of the misorientation for symmetric tilt grain boundaries in bcc iron. The two curves correspond to [100] and [110] interface normals. The temperature is $T_M-T=80\, \mathrm{K}$.
% temperature with the Thermocalc value of L: 107K. The "original" value was 80 K.
}
% data is in /Users/Spatschek/Documents/Physics/Ginzburg-Landau/bcc2D/ReadShockleyFailure
\label{RSfailure}
\end{center}
\end{figure}
Starting from a dense-packed configuration (see left panel of Fig.~\ref{rotlim::fig3}; inclination $\phi=0$) the misorientation is increased, and we see that the grain boundary energy increases monotonically.
It therefore does not reflect the proper sixfold symmetry which would imply that $\gamma_{gb}$ goes to zero for $\theta=60^\circ$.
Conversely, starting from the $\phi=30^\circ$ incliniation, the grain boundary does not ``heal'' if the dense-packed configuration is reached. A similar behavior is observed for bcc iron, where the amplitude equations do not obey the correct cubic symmetry, see Fig.~\ref{RSfailure}. For the reasons explained in section I, the amplitude equations are strictly valid only in the limit of small misorientations. However, for both hexagonal and bcc crystals the predictions remain approximately valid over roughly half the complete range allowed by the full crystal symmetry, e.g. $\gamma_{gb}$ is approximately valid between $0^\circ$ and $30^\circ$ for the $\phi=0^0$ inclination and $30^\circ$ and $60^\circ$ for the other $\phi=30^0$ inclination (and similarly for bcc on either side of $45^\circ$.)   A more detailed analysis for the simplest case of a smectic crystal is given in Appendix \ref{smectic}.

Fig.~\ref{RS::fig1} shows the grain boundary energy for small misorientations at a symmetrical grain boundary for the dense-packed crystal surfaces.
\begin{figure}
\begin{center}
\includegraphics[width=8.5cm]{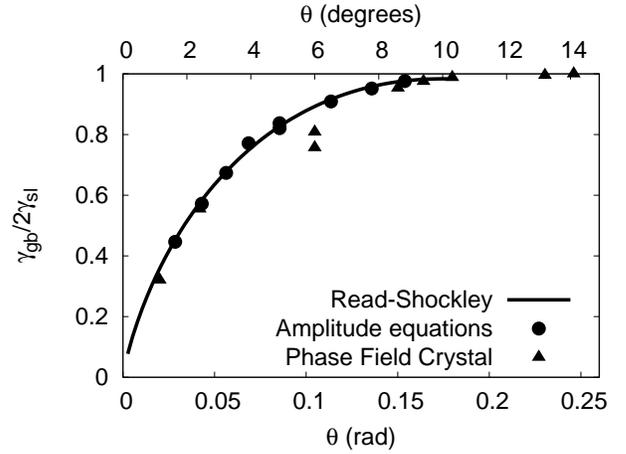}
\caption{Grain boundary energy as a function of misorientation for symmetrical tilt boundaries in a hexagonal crystal exactly at the melting point ($T=T_M$), normalized to twice the solid-liquid interfacial energy. The amplitude equation results are in very good quantitative agreement with the PFC results from Ref.~\onlinecite{Mellenthin08} for the same parameter $\epsilon_{2D}=0.1$. The boundary premelts for $\theta$ larger than $\theta_c\approx 10^\circ$.}
\label{RS::fig1}
\end{center}
\end{figure}
In contrast to Fig.~\ref{rotlim::fig3}, the temperature here is equal to the melting temperature.  
Above a critical misorientation $\theta_c\approx 10^\circ$, where $\gamma_{gb}\approx 2\gamma_{sl}$, the grains premelt, and the thickness of the melt layer diverges logarithmically as the melting point is approached from below.
The obtained data coincides well with the PFC simulations and a Read-Shockley fit, where the dislocation core radius is the only adjustable parameter.
A more detailed investigation of grain boundary premelting in the context of the ampltiude equations will be discussed elsewhere.

\subsection{Shear-induced grain boundary coupling and sliding}

\begin{figure}
\begin{center}
\includegraphics[width=7cm]{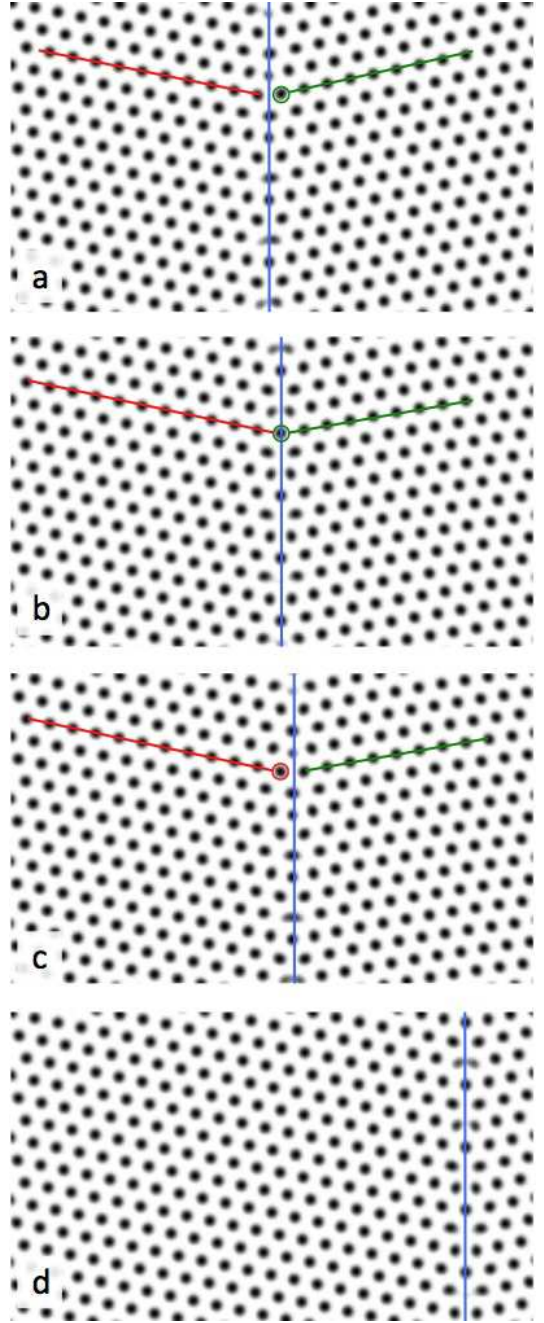}
\caption{Shear-induced coupled motion of two grains. The left grain is fixed and the right one is slowly pulled downwards. The blue line shows the location of the grain boundary, the red and green lines the crystallographic planes. The atom which is marked by the green circle reaches the atomic plane of the left crystal between (b) and (c) and attaches to the left crystal (red circle). Through this mechanism, the grain boundary moves perpendicular to the pulling direction.
Parameters are (in the notation of Appendix \ref{hex}) $\epspfc=0.1$, $K\epspfc=1$, with equal kinetic coefficients in the solid and the melt. The real space discretization is $\Delta x=0.25$, timestep $\Delta t=0.1$, and the misorientation between the two crystals at the symmetric tilt is $23.28^\circ$. The shear rate is $\dot{\epsilon}_{xy} = -10^{-4}$, and the dimensionless undercooling $\tilde{T}=-0.1$. The snapeshots are taken from a to d at times $0.1$, $9.0$, $12.5$ and $51.4$, respectively.
}
\label{coupling::fig1}
\end{center}
\end{figure}

\begin{figure}
\begin{center}
\includegraphics[width=7cm]{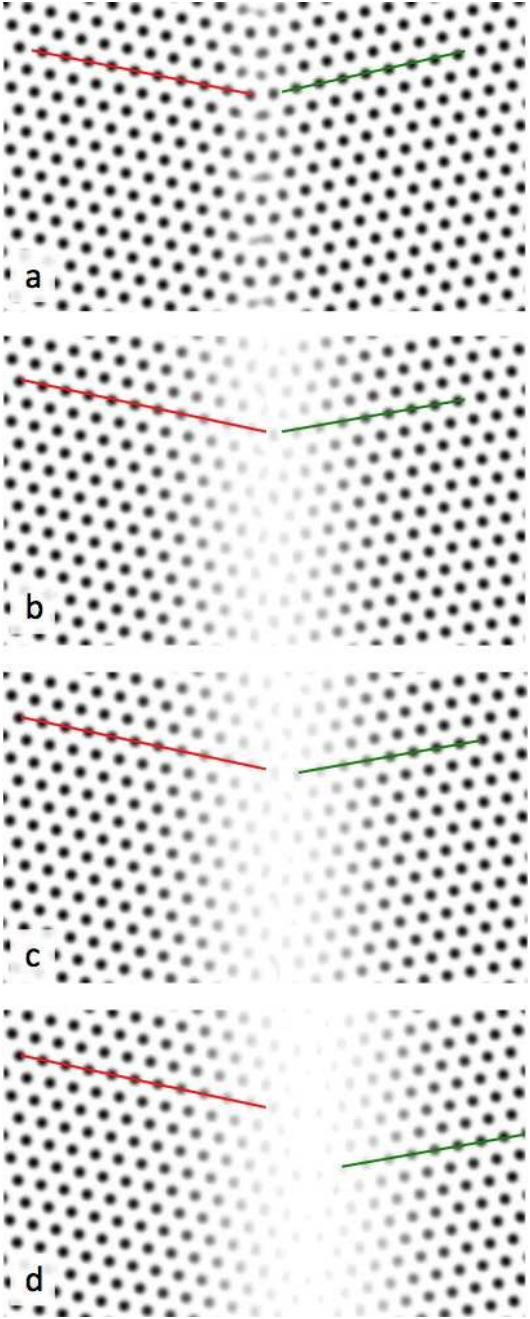}
\caption{Sliding of two grains. The left grain is fixed, the right one is slowly pulled downwards. The red and green mark lines of the crystallographic planes.
First, the system quickly equilibrates to a finite separation of the grains. Notice that the shear stress that builds up while the two grains are still connected also favors melting. Afterwards, the separation and the interface positions do not change any more and only the right grain slides downward.
%The misorientation and shear rate is the same as in Fig.~\ref{coupling::fig1}, but the temperature is closer to the melting point.
%{\bf I should eventually take some pictures from a later stage. Also, there seems to be an asymmetry, maybe I should do another run in a larger system.}
The parameters are the same as in Fig.~\ref{coupling::fig1}, with the exception of the much smaller dimensionless undercooling $\tilde{T}=-0.00012$.
}
\label{coupling::fig2}
\end{center}
\end{figure}

If a bicrystal is sheared in the direction parallel to a grain boundary, it migrates in a direction normal to the
grain boundary plane for low temperatures \cite{CahTay04, Cahetal06a, Cahetal06b, Mishetal2010}, and this effect is contained in the amplitude equation formulation.
%The present theory is particularly suited to model this effect on a continuum level, since it contains the necessary atomic resolution.
In Fig.~\ref{coupling::fig1}, the right crystal is sheared downwards.
Motion of the grain boundary by one lattice unit takes place during the time that an atom of the sheared right crystal needs to move until it matches the lattice of the left grain.
Then the grain boundary shifts in normal direction with velocity $v_\perp=v_s/[2\tan(\theta/2)]$ for $-\pi/6<\theta<\pi/6$, where $v_s$ is the sliding velocity.
This confirms the geometrical model of coupling in Refs. \onlinecite{CahTay04, Cahetal06a, Cahetal06b}, here applied to the hexagonal crystal symmetry.
The results in Fig. \ref{coupling::fig2} also show that the amplitude equation approach reproduces the transition from 
coupled motion to sliding\cite{Cahetal06b}, where the latter is favored close enough to the melting point. A more detailed study of this transition in an amplitude equation framework is currently in progress.

For the simulation we use a real space implementation, since we do not have periodic boundary conditions in the direction of the grain boundary normal.
At the left boundary $x=0$ we keep all amplitudes fixed in time
\begin{eqnarray*}
A_1(x=0) &=& A_s^0 \exp \frac{i\khnd{1}\mathbf{M}_l\Rv}{\epspfc^{1/2}}, \\
A_2(x=0) &=& -A_s^0 \exp \frac{i\khnd{2}\mathbf{M}_l\Rv}{\epspfc^{1/2}}, \\
A_3(x=0) &=& A_s^0 \exp \frac{i\khnd{3}\mathbf{M}_l\Rv}{\epspfc^{1/2}},
\end{eqnarray*}
with $\mathbf{M}_l=\mathbf{R}_l-\mathbf{I}$ and the two-dimensional rotation matrix $\mathbf{R}_l$ for the rotation of the left grain, compare also to equation (\ref{box::eq2}) and the notation introduced in Section \ref{box}.
The boundary conditions for the right grain at $x=X$ involve not only a rotation with $\mathbf{R}_r=\mathbf{R}_l^\dagger$ in the opposite direction, but also a time-dependent displacement:
\begin{eqnarray*}
A_1(x=X, t) &=& A_s^0 \exp \frac{i\khnd{1}\mathbf{M}_r\Rv}{\epspfc^{1/2}} \exp[-i \khnd{1} \mathbf{R}_r^\dagger \vec{u}(X, t)], \\
A_2(x=X, t) &=& -A_s^0 \exp \frac{i\khnd{2}\mathbf{M}_r\Rv}{\epspfc^{1/2}} \exp[-i \khnd{2} \mathbf{R}_r^\dagger \vec{u}(X, t)], \\
A_3(x=X, t) &=& A_s^0 \exp \frac{i\khnd{3}\mathbf{M}_r\Rv}{\epspfc^{1/2}} \exp[-i \khnd{3} \mathbf{R}_r^\dagger \vec{u}(X, t)],
\end{eqnarray*}
where the displacement vector $\vec{u}$ has components $u_x=0$ and $u_y=2\dot{\epsilon}_{xy}^0 t X$, with the strain rate $\dot{\epsilon}_{xy}^0$ (defined here on the ``slow'' scale).
Notice in particular that the elastic deformation factor involves the {\em rotated} principal reciprocal vectors $\mathbf{R}_r\khn{j}$.

%%%%%%%%%%%%%%%%%%%%%%%%%%%%%%%%%%%%%%%%%%
\begin{acknowledgments}

This work was supported by DOE through grant DE-FG02-07ER46400 and the Computational Materials Science Network program. R.S. also acknowledges financial support for the later part of this work of the German DFG grant SPP 1296 and from the industrial sponsors of ICAMS, ThyssenKrupp Steel AG, Salzgitter Mannesmann Forschung GmbH, Robert Bosch GmbH, Bayer Materials Science AG, Bayer Technology Services GmbH, Benteler AG and the state of North-Rhine-Westphalia.

\end{acknowledgments}

%%%%%%%%%%%%%%%%%%%%%%%%%%%%%%%%%%%%%%%%%%
%%%%%%%%%%%%%%%%%%%%%%%%%%%%%%%%%%%%%%%%%%
\appendix

%%%%%%%%%%%%%%%%%%%%%%%%%%%%%%%%%%%%%%%%%%
\section{Phase field crystal}
\label{hex}

For completeness, we derive in this appendix the amplitude equations from the phase field crystal model 
for two-dimensional hexagonal crystals using the same NWS type multiscale expansion as for bcc \cite{Wu07}.

We start from the dimensionless free energy functional
\begin{equation} \label{hex::eq1}
{\cal F} = \int d\rv \left( \frac{\psi}{2} \left[ -\epspfc + (\nabla^2 + 1)^2 \right] \psi + \frac{1}{4}\psi^4  \right)
\end{equation}
The parameter $\epspfc$ plays the same role as the scale separation parameter $\epsnew$ as defined in section \ref{dw}.
Equilibrium requires that the chemical potential
\begin{equation} \label{hex::eq2}
\mu_E = \frac{\delta {\cal F}}{\delta \psi} = -\epspfc\psi + (\nabla^2 + 1)^2\psi + \psi^3
\end{equation}
is spatially constant.
The density $\psi$ is constant in the liquid, $\psi=\bar{\psi}_l$, thus we get the free energy density
\begin{equation} \label{hex::eq3}
\bar{f}_l = -(\epspfc-1)\frac{\bar{\psi}_l^2}{2} + \frac{\bar{\psi}_l^4}{4}.
\end{equation}
We use a one-mode approximation for the solid,
\begin{equation} \label{hex::eq4}
\psi(\rv) = \bar{\psi}_s + \sum_{j=1}^N \Ajpfc \exp(i\kvj\cdot\rv)
\end{equation}
with the following set of $N=6$ normalized principal reciprocal lattice vectors:
\begin{eqnarray}
\khn{1} = \frac{1}{2} \left(
\begin{array}{c}
-\sqrt{3} \\
-1
\end{array}
\right),
& 
\khn{2} = \left(
\begin{array}{c}
0 \\
1
\end{array}
\right),
&
\khn{3} = \frac{1}{2} \left(
\begin{array}{c}
\sqrt{3} \\
-1
\end{array}
\right), \nonumber \\
\khn{\bar{1}} = \frac{1}{2} \left(
\begin{array}{c}
\sqrt{3} \\
1
\end{array}
\right),
& 
\khn{\bar{2}} = \left(
\begin{array}{c}
0 \\
-1
\end{array}
\right),
&
\khn{\bar{3}} = \frac{1}{2} \left(
\begin{array}{c}
-\sqrt{3} \\
1
\end{array}
\right). \label{hex::eq5}
\end{eqnarray}
The k-vectors used in the expansion above are a multiple of these vectors, and their length is determined by a free energy minimization below;
in fact, we will obtain then $\kvj=\khj$.
The amplitudes are real with
\begin{equation} \label{hex::eq6}
A_1 = A_{\bar{1}} = -A_2 = -A_{\bar{2}} = A_3 = A_{\bar{3}}= A_s,
\end{equation}
thus leading to \cite{Eldetal04,Mellenthin08}
\begin{equation} \label{hex::eq7}
\psi_s = \bar{\psi}_s + 4A_s \left[ \cos qx \cos \frac{qy}{\sqrt{3}} - \frac{1}{2}\cos\frac{2qy}{\sqrt{3}} \right],
\end{equation}
where the factor $q$ is introduced to find the correct length of the reciprocal lattice vectors, which corresponds to the atomic spacing.
We can then calculate the free energy (per unit cell) and minimize it with respect to $q$ and $A_s$.
This gives \cite{Mellenthin08}
\begin{equation} \label{hex::eq8}
A_s = \frac{1}{5} \left( \bar{\psi}_s \pm \frac{1}{3} \sqrt{15\epspfc - 36\bar{\psi}_s^2} \right)
\end{equation}
where the $\pm$ sign is for positive and negative $\bar{\psi}_s$ respectively;
in accordance with Ref.~\onlinecite{Mellenthin08} we pick the negative branch.
Furthermore, we obtain $q=\sqrt{3}/2$ (thus $\kvj=\khj$).
Then the average free energy density in the solid is
\begin{eqnarray}
\bar{f}_s &=& -\frac{1}{10} \epspfc^2 - \frac{13}{500}\bar{\psi}_s^4 + \frac{1}{2}\bar{\psi}_s^2 \left( 1- \frac{7}{25}\epspfc \right) \nonumber \\
&+& \frac{4}{25}\bar{\psi}_s \sqrt{-36\bar{\psi}_s^2 + 15\epspfc} \left( -\frac{4}{5} \bar{\psi}_s^2 + \frac{1}{3}\epspfc \right). \label{hex::eq9}
\end{eqnarray}
Coexistence between solid and liquid demands the equality of the chemical potentials, $\mu\equiv\mu_{s/l}=\partial \bar{f}_{s/l}/\partial\bar{\psi}_{s/l}$, and the grand potentials, $\omega\equiv \omega_{s/l}=\bar{f}_{s/l}-\bar{\psi}_{s/l}\mu$, of the two phases.

In the spirit of a multiscale expansion we write the average densities as
\begin{eqnarray}
\bar{\psi}_s &=& \psi_{s0}\epspfc^{1/2} + \psi_{s1}\epspfc + \psi_{s2}\epspfc^{3/2} + \ldots \label{hex::eq10} \\
\bar{\psi}_l &=& \psi_{l0}\epspfc^{1/2} + \psi_{l1}\epspfc + \psi_{l2}\epspfc^{3/2} + \ldots. \label{hex::eq11}
\end{eqnarray}
Up to the order $\epspfc$ the chemical potential difference is
\begin{equation} \label{hex::eq12}
\mu_s - \mu_l = (\psi_{s0}-\psi_{l0})\epspfc^{1/2} + (\psi_{s1}-\psi_{l1})\epspfc + {\cal O}(\epspfc^{3/2}),
\end{equation}
which implies
\begin{eqnarray}
\psi_{s0}&=&\psi_{l0}, \label{hex::eq13} \\
\psi_{s1}&=&\psi_{l1}. \label{hex::eq14}
\end{eqnarray}
From the next order term of the chemical potential balance, ${\cal O}(\epspfc^{3/2})$, we get a relation between $\psi_{l2}$ and $\psi_{s2}$,
\begin{eqnarray*}
\psi_{l2} =& \Big( & -125 \psi_{s2} \sqrt{-36\psi_{s0}^{2}+15} \\
&&- 90\sqrt{-36\psi_{s0}^{2}+15}\psi_{s0} +1200\psi_{s0}^{2}-100 \\
&&+138\sqrt{-36\psi_{s0}^{2}+15} \psi_{s0}^{3} -2304\psi_{s0}^{4}\Big) \\
&/&\Big(-125\sqrt{-36\psi_{s0}^{2}+15}\Big).
\end{eqnarray*}
Up to the order $\epspfc^{3/2}$ the difference between the grand potentials vanishes, and from the order $\epspfc^2$ we get the result
\begin{equation} \label{hex::eq15}
\psi_{s0} = -\frac{\sqrt{555}}{37} \approx -0.6367
\end{equation}
(there is also another solution, $1/\sqrt{3}$, which we drop, since we concentrate here on the negative branch).
Similarly, from the chemical potential balance at order $\epspfc^2$ we obtain $\psi_{l1}=\psi_{s1}=0$.

Beyond the thermodynamical analysis above, which deals only with the spatially averaged quantities, we consider now explicitly the spatial oscillations of the density.
The key is the separation of slow variables, denoted by capital letters $\Rv$, and fast variables, $\rv$.
They are related by the expansion parameter $\epspfc$, $\Rv= \epspfc^{1/2} \rv$.
This translates also to gradients, where we introduce two gradients, $\snabla$ and $\lnabla$, where the first operator acts only on fast, the second on slow variables, thus $\nabla \rightarrow \snabla + \epspfc^{1/2} \lnabla$.
We therefore obtain the transformation rule
\begin{eqnarray*}
(\nabla^2+1)^2 &\rightarrow& (\snabla^2+1)^2 + 4\epspfc^{1/2}(\snabla^2+1) \snabla\cdot\lnabla \\
&& + 2\epspfc \left[ (\snabla^2+1)\lnabla^2 + 2(\snabla\cdot\lnabla)^2 \right],
\end{eqnarray*}
where we skipped terms of order $\epspfc^{3/2}$ and higher.

We expand the field according to
\begin{equation} \label{hex::eq16}
\psi(\rv) = \psi_0(\rv)\epspfc^{1/2} + \psi_1(\rv)\epspfc + \psi_2(\rv)\epspfc^{3/2} +\ldots.
\end{equation}
Using the expansion for the averaged densities Eqs.~(\ref{hex::eq10}) and (\ref{hex::eq11}), the multiscale phase equilibrium version of the PFC equation (\ref{hex::eq2}) is
\[
-\epspfc\psi + (\nabla^2+1)^2\psi + \psi^3 = \psi_{s0} \epspfc^{1/2} + (\psi_{l2}-\psi_{s0}+\psi_{s0}^3)\epspfc^{3/2} + \ldots,
\]
which is of course the same as for bcc \cite{Wu07}, but with a different value $\psi_{s0}$.

At order $\epspfc^{1/2}$ we obtain the equation
\begin{equation} \label{hex::eq17}
(\snabla^2+1)^2 \psi_0(\rv) = \psi_{s0},
\end{equation}
which is solved by
\begin{equation} \label{hex::eq18}
\psi_0(\rv) = \psi_{s0} + \sum_{j=1}^N A_j^0(\Rv) e^{i\khj\cdot \rv}.
\end{equation}
At order $\epspfc$, we have
\begin{equation} \label{hex::eq19}
(\snabla^2+1)^2 \psi_1(\rv) = 0,
\end{equation}
with the solution
\begin{equation} \label{hex::eq20}
\psi_1(\rv) = \sum_{j=1}^N A_j^1(\Rv) e^{i\khj\cdot \rv}.
\end{equation}

At order $\epspfc^{3/2}$ we have for the first time also gradients with respect to the slow variables,
\begin{equation} \label{hex::eq21}
(\snabla^2+1)^2 \psi_2 -\psi_0 + 4(\snabla\cdot\lnabla)^2\psi_0 + \psi_0^3 = \psi_{l2} -\psi_{s0} + \psi_{s0}^3.
\end{equation}
All secular terms proportional to $\exp(i\khj\cdot\rv)$ must balance each other, thus we get e.g. for the prefactor in front of $\exp(i\khn{1}\cdot\rv)$
\begin{eqnarray*}
&& -A_1^0  - 4(\khn{1}\cdot \lnabla)^2 A_1^0 + 6\psi_{s0} A_2^{0*} A_3^{0*} + 3\psi_{s0}^2 A_1^0 \\
&& + 6[ A_2^0 A_2^{0*} + A_3^0 A_3^{0*} ] A_1^0 + 3 A_1^0 A_1^{0*} A_1^0 = 0.
\end{eqnarray*}
For a pure solid, we can find solution with constant real amplitudes $A_j^0$, with $A_1^0=A_3^0 = A_s^0 := -4/\sqrt{555}$, and $A_2^0=-A_s^0$;
this solution minimizes the free energy given below with $F=0$.

The above equation (and its counterpart for the prefactors of the other exponential terms) can be obtained variationally from the following free energy:
\begin{eqnarray}
F &=& F_{2D}^0 \int d\Rv \Bigg\{ 4 \sum_{j=1}^{N/2} \left| (\khj\cdot\lnabla) A_j^0 \right|^2 \nonumber \\
&& + (3\psi_{s0}^2-1) \sum_{j=1}^{N/2} A_j^0 A_j^{0*} \nonumber \\
&&+ 6\psi_{s0} (A_1^{0*} A_2^{0*} A_3^{0*} + A_1^{0} A_2^{0} A_3^{0}) \nonumber \\
&& + 3\left(\sum_{j=1}^{N/2} A_j^0 A_j^{0*} \right)^2 - \frac{3}{2} \sum_{j=1}^{N/2} |A_j^0|^4 \Bigg\}, \label{hex::eq22}
\end{eqnarray}
%\comment{What is the proper prefactor of the free energy functional? I took now $\epspfc^{1/2}$.}
where we treat real and imaginary of the complex fields $A_j^0$ as independent for the variation.
The prefactor of the functional, $F_{2D}^0$, is not determined by the equilibrium condition, and will be determined by consideration of elastic deformations below.
The summation up to $N/2$ means that we sum over $1, 2, 3$.
Notice that the same expression holds for bcc up to quadratic order, apart from the fact the $N=12$ there for the different set of principal reciprocal lattice vectors.
The cubic and quartic terms all satisfy the conditions that only vectors, which form closed polygons, contribute;
this corresponds to the appearance of a $\delta$-function when the fast oscillations are integrated out.
%Here, these terms are determined from the expansion of underlying PFC model, but they can also be assumed using an equal weight ansatz for equivalent terms (does this give different weight factors?).
%\comment{Do we get the same result with the equal weight ansatz?}

The generalization to a rotational invariant form is via the replacement
\begin{equation}
\kvj\cdot\lnabla \longrightarrow \Boxpfc_j = \kvj\cdot\lnabla - \frac{i\epspfc^{1/2}}{2} \lnabla^2.
\end{equation}

The equilibrium conditions, that lead to the cancellation of the secular terms above is
\begin{equation} \label{hex::eq23}
\frac{\delta F}{\delta A_j^{0*}} = 0.
\end{equation}
We can formulate a dynamical form of these equations by
\begin{equation} \label{hex::eq24}
\frac{\partial A_j^0}{\partial t} = - K \frac{\delta F}{\delta A_j^{0*}},
\end{equation}
with a kinetic coefficient $K$.
Notice that the free energy decays monotonically with these evolution equations for the complex fields, and we therefore finally reach an equilibrium state.

Analogous to the general expression for the elastic constants (\ref{elast::eq12}) we obtain 
\begin{equation} \label{hex::eq25}
c_{ijkl} = F_{2D}^0 (A_s^0)^2 \times
\left\{
\begin{array}{cc}
9 & \mbox{if } i=j=k=l \\
3 & \mbox{for two distinct pairs of indices} \\
0 & \mbox{else}
\end{array}
\right.
\end{equation}
We note that these expressions are defined on the ``fast scale'' $\rv$.
As expected, this case corresponds to isotropic elasticity, and the usual Lam\'e coefficient and shear modulus are
\begin{eqnarray}
\lambda &=& 3 F_{2D}^0 (A_s^0)^2 \label{hex::eq26}\\
\mu &=& 3 F_{2D}^0 (A_s^0)^2. \label{hex::eq27}
\end{eqnarray}
This corresponds to a (three-dimensional) Poisson ratio of $\nu=1/4$.
%\comment{The elastic constants contain here one factor $\epspfc$ from the 2D conversion to the fast scale, the other $\epspfc^{1/2}$ comes from the above free energy expression. So if this needs to be adopted, we also have to do it here.}

Finally, this allows to determine the energy scale $F_{2D}^0$ by calculation of the elastic energy of a deformed solid, where all amplitudes are equal to $A_s^0$:
In the energy density of the phase field crystal model, e.g.~a (small) strain $\epsilon_{xx}$ leads to an increase of the free energy density by $\bar{f}_s-\bar{f}_l = (24/185) \epsilon_{xx}^2 \epspfc$ to lowest order in $\epspfc$, if all other strain components vanish.
On the other hand, the same free energy density change (\ref{hex::eq22}) for the amplitude equations, written of the fast scale, is $f_{el} = \sigma_{ij}\epsilon_{ij}/2 = (\lambda +\mu/2) \epsilon_{xx}^2$ with the elastic constants given in Eqs.~(\ref{hex::eq26}) and (\ref{hex::eq27}).
Thus we have $f_{el} = (24/185) F_{2D}^0 \epsilon_{xx}^2$, and therefore
\begin{equation}
F_{2D}^0 = \epspfc.
\end{equation}

A deviation from the melting temperature and the coupling to thermodynamic alloy models is achieved via an additional free energy term
\begin{equation}
F_T = \int d\Rv\, \phi \Tpfc
\end{equation}
with a dimensionless temperature deviation $\Tpfc$ from the melting point;
the interpolating ``phase field'' is defined as in Eqs. (\ref{dw::eq5b}) and (\ref{dw::eq5d}).

We can eliminate the phase field crystal parameters by rescaling the equations using
\begin{equation} \label{hex::eq28}
\Ajpfcnew = A_j^0/A_s
\end{equation}
and introduction of another (small) dimensionless parameter
\begin{equation} \label{hex::eq29}
\epspfcnew = \frac{2}{3} \frac{1}{3\psi_{s0}^2-1} \epspfc.
\end{equation}
Similarly, the length scales are scaled with this new parameter, $\tilde{X}=\epspfcnew^{1/2} x$ and
\begin{equation} \label{hex::eq30}
\Boxpfcnew_j = \kvj\cdot\nnabla - \frac{i\epspfcnew^{1/2}}{2} \nnabla^2.
\end{equation}
Then the free energy becomes
\begin{eqnarray}
F  &=& \tilde{F}_{2D}^0 \int d\tilde{R} \Bigg\{ \sum_{j=1}^{N/2} \left| \Boxpfcnew_j \Ajpfcnew \right|^2 \nonumber + \frac{1}{6} \sum_{j=1}^{N/2} \Ajpfcnew \Ajpfcnew^* \nonumber \\
&&+ \frac{1}{2} (\Apfcnew_1^{*} \Apfcnew_2^{*} \Apfcnew_3^{*} + \Apfcnew_1 \Apfcnew_2 \Apfcnew_3) \nonumber \\
&& + \frac{1}{15} \left(\sum_{j=1}^{N/2} \Apfcnew_j \Apfcnew_j^{*} \right)^2 - \frac{1}{30} \sum_{j=1}^{N/2} |\Apfcnew_j|^4 \Bigg\}\label{hex::eq31}
\end{eqnarray}
with
\begin{equation} \label{hex::eq32}
\tilde{F}_{2D}^0 = 4 A_s^2F_{2D}^0 = \frac{256}{6845} \epspfcnew. 
\end{equation}

%A deviation from the melting temperature and the coupling to thermodynamic alloy models is achieved via an additional free energy term
%\begin{equation}
%F_c +F_T = \int d\Rv \left[ \Tpfc\phi + \phi\depfc c + \bpfc (c\ln c - c) \right]
%\end{equation}
%with a dimensionless temperature deviation $\Tpfc$ from the melting point, an entropic coefficient $\bpfc$ and an energy penalty for impurities in the solid $\depfc$, similar to Eq (\ref{alloy::eq1});
%the interpolating ``phase field'' is defined as in Eqs. (\ref{dw::eq5b}) and (\ref{dw::eq5d}).
%For differently fast diffusion in solid and liquid we interpolate between the solid and the liquid diffusion coefficients, $D_s$ and $D_l$, respectively, via
%\begin{equation}
%D = \phi D_s + (1-\phi) D_l.
%\end{equation}
%The impurity diffusion is then described by
%\begin{eqnarray}
%\frac{\partial c}{\partial t} &=& \lnabla\cdot \left[ D c \lnabla \frac{\delta F}{\delta c} \right] \nonumber \\
%&=& \lnabla\cdot \left[ D c \depfc \lnabla\phi + D \bpfc \lnabla c \right]. 
%\end{eqnarray}
%This model corresponds to a phase diagram with straight solidus and liquidus line, with partition coefficient $c_s/c_l=\kpfc=\exp(-\depfc/\bpfc)$.
%The liquidus line is given by
%\begin{equation}
%\Tpfc = -\bpfc(1-\kpfc) c_l.
%\end{equation}
%To describe the lattice expansion due to impurities, we replace
%\begin{equation}
%\Boxpfc_j \rightarrow \Boxpfc_j + i\alpha c.
%\end{equation}

%%%%%%%%%%%%%%%%%%%%%%%%%%%%%%%%%%%%%%%%%%
\section{Breakdown of rotational invariance}
\label{smectic}

To illustrate the breakdown of rotational invariance, we consider the simple case of a smectic crystal, which is described by only two antiparallel principal reciprocal lattice vectors and only one complex amplitude $u$.
%\begin{equation}
%\khn{1} = \left( \begin{array}{c} 1 \\ 0 \\ 0 \end{array} \right), \qquad \khn{2} = \left( \begin{array}{c} -1 \\ 0 \\ 0 \end{array} \right).
%\end{equation}
A pure crystal is then described by the amplitude $u(X)=1$ and the density variation
\begin{equation} \label{smectic::eq1}
\delta n(x) = u \exp(i x) + u^* \exp(-ix)
\end{equation}
with $x=X/\epsilon^{1/2}$.
If the crystal is rotated by $180^\circ$, i.e.~$u(X)=\exp(-2ix)$, it recovers its original state according to Eq.~(\ref{smectic::eq2}) with the same density $\delta n$.
Due to the high rotation angle, the spacing between the beats in the amplitude is half the lattice spacing, thus a fine discretization is necessary for a numerical implementation.
Notice that both states are purely one-dimensional.

This system is described through a (dimensionless) one-dimensional free energy functional
\begin{equation} \label{smectic::eq2}
F_{\mathrm{smectic}} = \int dX \left[ f_k + f_p + f_T \right] 
\end{equation}
with $f_k = |\Box u|^2$, $\Box = \partial_X - i\epsilon^{1/2} \partial_X^2/2$, $f_p = |u|^2 (1-|u|^2)^2$ and $f_T= L(T-T_M)h(|u|^2)/T_M$.
In equilibrium, a one-dimensional solution is therefore described by the ordinary differential equation
\begin{equation}
u'' - i\epsilon^{1/2} u''' - \epsilon u''''/4 = \frac{df_p}{du} + \frac{df_T}{du},
\end{equation}
where the prime $'$ denotes differentiation with respect to $X$.
Obviously, both above one-dimensional states satisfy this equation, therefore the rotational invariance of the bulk states holds.

\begin{figure}
\begin{center}
\includegraphics[width=9cm]{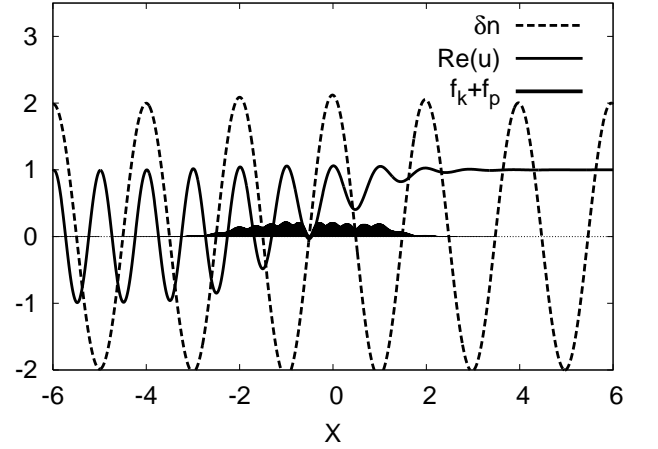}
\caption{Density variation $\delta n(X)$ and real part of the amplitude $u(X)$ of a smectic crystal that consists of one ``grain'' that is not rotated ($X>0$) and one ``grain'' that is rotated by $180^\circ$ ($X<0$).
The density seems to indicate a perfectly healed crystal, but nevertheless an interface energy is associated with the interface, leading to a finite energy density $f_k+f_p$ in the interface region and therefore a nonvanishing grain boundary energy (solid black area).
The parameters used here are $\epsilon=0.1$ and $L(T-T_M)/T_M=-0.1$.}
\label{smectic::fig1}
\end{center}
\end{figure}

\begin{figure}
\begin{center}
\includegraphics[width=9cm]{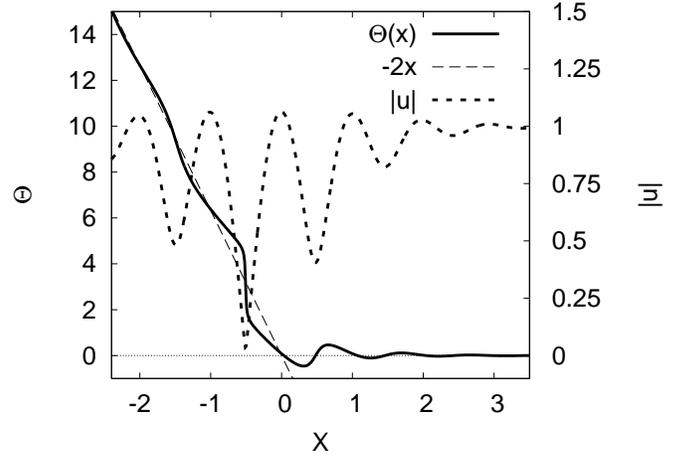}
\caption{Phase and modulus of the amplitude $u$ in the interface region, corresponding to the results and parameters in Fig.~\ref{smectic::fig1}.}
\label{smectic::fig2}
\end{center}
\end{figure}

Although both amplitudes describe the same density, a defect free interface cannot be formed between them.
In particular, for $T=T_M$  a ``grain boundary'' between the state $u=\exp(-2ix)$ for $x\to -\infty$ and $u=1$ for $x\to\infty$ would premelt, and a melt layer forms between the two ``grains''.
In an undercooled state, $T<T_M$, Fig.~\ref{smectic::fig1} shows the reconstructed equilibrium density, which seems to be defect-free.
In the same figure, also the corresponding amplitude and the free energy density (without the contribution from the thermal tilt) are shown.
At the interface, the phase $\Theta(X)$ of the amplitude, $u=|u|\exp(i\Theta)$, changes smoothly, see Fig.~\ref{smectic::fig2}.
Obviously, a finite interface forms between the two grains, and it is accompanied by a finite energy density, thus a spurious finite grain boundary energy is found.

%%%%%%%%%%%%%%%%%%%%%%%%%%%%%%%%%%%%%%%%%%
\section{Periodic boundary conditions}
\label{periodic}

For purposes of numerical modeling using spectral methods \cite{Mellenthin08}, periodic boundary conditions are advantageous.
This becomes restrictive here, since we need the periodicity for each order parameter.
For the simplest case of a liquid or a solid that is not rotated with respect to the ``natural'' orientation of the set of principal reciprocal lattice vectors, the amplitudes are constant in each phase, and therefore no periodicity constraints arise.
This is also true for coexistence of solid and liquid.

The situation becomes more complex if rotated crystals are involved, because then the density wave amplitudes acquire a periodic modulation.
We look at the 2D hexagonal system with size $X\times Y$ first and use the notation of Appendix \ref{hex}.
Let us consider the situation of a single solid phase, then the amplitudes of the rotated crystal are given by
\begin{equation} \label{periodic::eq1}
\Ajpfc^0(\rv) = A_s^0 \exp(i\khjt\Mmatrix\rv)
\end{equation}
Periodicity requires therefore at the horizontal boundary at $x=0$ and $x=X$
\begin{equation} \label{periodic::eq2}
\exp\left[i\khjt\Mmatrix \left( \begin{array}{c} 0 \\ y \end{array} \right)\right] = \exp\left[i\khjt\Mmatrix \left( \begin{array}{c} X \\ y \end{array} \right)\right]
\end{equation}
for $j=1\ldots N$ and arbitrary vertical coordinate $y$.
Hence,
\begin{equation} \label{periodic::eq3}
\khjt\Mmatrix \left( \begin{array}{c} X \\ 0 \end{array} \right) = 2\pi n_j
\end{equation}
with integer numbers $n_j$.
Summation over the first $N/2=3$ principal reciprocal lattice vectors (see Eq.~(\ref{hex::eq5})), which form a closed triangle, therefore gives 
\begin{equation} \label{periodic::eq4}
n_1+n_2+n_3=0.
\end{equation}
(Since the fields that are associated to the other principal reciprocal lattice vectors are complex conjugate to the previous, their periodicity does not give additional conditions).
From the definition of the rotation matrix Eq.~(\ref{box::eq3}) we therefore get the two conditions
\begin{eqnarray}
-\frac{\sqrt{3}}{2} (\cos\theta -1) + \frac{1}{2}\sin\theta &=& \frac{2\pi n_1}{X}, \label{periodic::eq5} \\
-\sin\theta &=& \frac{2\pi n_2}{X}. \label{periodic::eq6}
\end{eqnarray}
From these two equations we obtain immediately
\begin{equation} \label{periodic::eq7}
-\sqrt{3}(\cos\theta-1) = \frac{2\pi}{X}(2n_1+n_2).
\end{equation} \label{periodic::eq8}
Subsequent division by Eq.~(\ref{periodic::eq6}) and use of trigonometric identities yields
\begin{equation} \label{periodic::eq9}
-\sqrt{3}\tan\frac{\theta}{2} = \frac{2n_1+n_2}{n_2},
\end{equation}
which defines a discrete set of admissible of lattice rotation angles $\theta$, given through the integer number $n_1, n_2$.
The corresponding system length is then
\begin{equation} \label{periodic::eq10}
X = \frac{2\pi n_2}{-\sin\theta},
\end{equation}
which obviously becomes large for low angle rotations.
The analogous expressions for periodicity in $y$ direction in a system of height $Y$ are
\begin{equation} \label{periodic::eq11}
\tan\frac{\theta}{2} = \sqrt{3}\frac{m_2}{2m_1+m_2},
\end{equation}
where $m_1, m_2$ are also integer numbers.
the system height is then
\begin{equation} \label{periodic::eq12}
Y = \frac{2\pi m_2}{\cos\theta -1}.
\end{equation}
Now, both conditions (\ref{periodic::eq9}) and (\ref{periodic::eq11}) must be satisfied, therefore giving
\begin{equation} \label{periodic::eq13}
-\frac{2n_1+n_2}{n_2} = 3\frac{m_2}{2m_1+m_2},
\end{equation}
for which integer solutions have to be found.
Since we are interested in finding small system sizes, i.e. small values of $n_2$ and $m_2$, we can choose $m_2=-1$ and $m_1<0$.
Using for example $n_1 = -m_1-2m_2$ and $n_2=2m_1+m_2$ therefore gives a solution of the above equation with the angle given by Eq.~(\ref{periodic::eq11}), and the system sizes follow from Eqs. (\ref{periodic::eq10}) and (\ref{periodic::eq12}).
For specific cases, also solutions with smaller systems sizes can be found.

Although the requirement of periodicity for a bcc crystal seems to be more stringent at a first glance due to the higher number of order parameters, this complexity is significantly reduced by the ability to form different sets of closed polygons of principal reciprocal lattice vectors.
First, we have now six conditions of the type (\ref{periodic::eq3}), but from the formation of closed triangles we get the integer relations
\begin{eqnarray}
-n_{011}+n_{101}-n_{1\bar{1}0} &=& 0, \label{periodic::eq14} \\
-n_{011}+n_{110}-n_{10\bar{1}} &=& 0, \label{periodic::eq15} \\
-n_{01\bar{1}} + n_{110} - n_{101} &=& 0, \label{periodic::eq16} \\
-n_{01\bar{1}} + n_{10\bar{1}} - n_{1\bar{1}0} &=& 0, \label{periodic::eq17}
\end{eqnarray}
but only three of them are independent.
Also, the conditions of closed quadrilaterals does not provide additional independent information.

The periodicity condition for $[011]$ is explicitly in analogy to Eq.~(\ref{periodic::eq5})
\begin{equation} \label{periodic::eq18}
-\frac{1}{\sqrt{2}} \sin\theta X = 2\pi n_{011}
\end{equation}
and similar for $[101]$ and $[110]$
\begin{eqnarray}
\frac{1}{\sqrt{2}} (\cos\theta -1) X &=& 2\pi n_{101}, \label{periodic::eq19} \\
\frac{1}{\sqrt{2}} (\cos\theta -1) X -\frac{1}{\sqrt{2}} \sin\theta\, X &=& 2\pi n_{110}, \label{periodic::eq20}
\end{eqnarray}
from which get the additional (independent) integer relation
\begin{equation} \label{periodic::eq21}
n_{110} - n_{101} - n_{011} = 0.
\end{equation}
Therefore, only two integer numbers can be chosen independently, e.g.~$n_{101}$ and $n_{011}$, and we finally arrive at the expressions
\begin{eqnarray}
\tan\frac{\theta}{2} &=& \frac{n_{101}}{n_{011}}, \label{periodic::eq22} \\
X &=& - \frac{2\sqrt{2}\pi n_{011}}{\sin\theta}. \label{periodic::eq23}
\end{eqnarray}
Similarly, periodicity in $y$ direction implies the conditions
\begin{eqnarray}
\tan\frac{\theta}{2} &=& -\frac{m_{011}}{m_{101}}, \label{periodic::eq24} \\
Y &=& \frac{2\sqrt{2}\pi m_{101}}{\sin\theta} \label{periodic::eq25}
\end{eqnarray}
with the only two independent integer numbers $m_{101}$ and $m_{011}$.
The equality of the angles according to Eqs.~(\ref{periodic::eq22}) and (\ref{periodic::eq24}) demands therefore integer solutions of the equation
\begin{equation} \label{periodic::eq26}
-\frac{m_{011}}{m_{101}} = \frac{n_{101}}{n_{011}}.
\end{equation}

If we assume translation invariance of the amplitudes in $z$ direction (i.e.~no periodicity condition), the problem becomes effectively two-dimensional.
Notice, however, that the reconstructed density waves still have a periodic modulation in that direction, and also the atoms are not bound to stay in the $xy$ plane.
In fact, all displacements $u_x(x,y)$, $u_y(x,y)$ and $u_z(x,y)$, that depend only on in-plane coordinates, can be described by a two-dimensional formulation of the amplitude equations.

For a discussion concerning periodicity in systems with a grain boundary we refer to Ref.~\onlinecite{Mellenthin08}.

%%%%%%%%%%%%%%%%%%%%%%%%%%%%%%%%%%%%%%%%%%

\end{document}